\documentclass[lettersize,journal]{IEEEtran}
\usepackage{amsmath,amsfonts}
\usepackage{algorithmic}
\usepackage{enumitem}
\usepackage[linesnumbered,ruled,vlined]{algorithm2e}
\usepackage{setspace}
\usepackage{array}
\usepackage{textcomp}
\usepackage{stfloats}
\usepackage{multirow,diagbox,tabularx,blindtext}
\usepackage{booktabs}
\usepackage{hyperref}
\usepackage{verbatim}
\usepackage{graphicx}
\usepackage{amssymb}
\usepackage{textcomp}
\usepackage{cite}
\usepackage{color}
\usepackage{pifont}
\usepackage{makecell}
\usepackage[T1]{fontenc}
\usepackage[scaled=0.81]{beramono}

\newcommand{\PP}[1]{
\vspace{2px}
\noindent{\bf \IfEndWith{#1}{.}{#1}{#1.}}
}

\usepackage{subfigure}
\usepackage[normal]{caption}
\begin{document}

\begin{sloppypar}

\title{\huge Rethinking Smart Contract Fuzzing: Fuzzing With Invocation Ordering and Important Branch Revisiting} 

\author{Zhenguang Liu, Peng Qian, Jiaxu Yang, Lingfeng Liu, Xiaojun Xu, Qinming He, and Xiaosong Zhang

\thanks{This work was supported by the National Key R\&D Program of China under Grant 2021YFB2700500, the Key R\&D Program of Zhejiang Province under Grant 2022C01086 and Grant 2021C01104, and by the Scientific Research Fund of Zhejiang Provincial Education Department under Grant Y202250832. \emph{(Corresponding author: Peng Qian.)} }       

\thanks{Zhenguang Liu, Peng Qian, and Qinming He are with Zhejiang University, Hangzhou 310058, China (e-mail: liuzhenguang2008@gmail.com; messi.qp711@gmail.com; hqm@zju.edu.cn). }

\thanks{Jiaxu Yang, Lingfeng Liu, and Xiaojun Xu are with School of Computer and Information Engineering, Zhejiang Gongshang University, Hangzhou 310018, China (e-mail: yjx.00@foxmail.com; liulingfengxx@gmail.com; xuxj2022@gmail.com).}

\thanks{Xiaosong Zhang is with the Center for Cyber Security, University of Electronic Science and Technology of China, Chengdu 611731, China (e-mail: johnsonzxs@uestc.edu.cn).}

}

\markboth{IEEE TRANSACTIONS ON INFORMATION FORENSICS AND SECURITY} 
{Liu \MakeLowercase{\textit{et al.}}: Rethinking Smart Contract Fuzzing: Fuzzing With Invocation Ordering and Important Branch Revisiting} 


\maketitle

\begin{abstract}
Blockchain smart contracts have given rise to a variety of interesting and compelling applications and emerged as a revolutionary force for the Internet. Smart contracts from various fields now hold over one trillion dollars worth of virtual coins, attracting numerous attacks. Quite a few practitioners have devoted themselves to developing tools for detecting bugs in smart contracts. One line of efforts revolve around static analysis techniques, which heavily suffer from high false positive rates.  Another line of works concentrate on fuzzing techniques. Unfortunately, current fuzzing approaches for smart contracts tend to conduct fuzzing starting from the initial state of the contract, which expends too much energy revolving around the initial state of the contract and thus is usually unable to unearth bugs triggered by other states. Moreover,  most existing methods treat each branch equally, failing to take care of the branches that are rare or more likely to possess bugs. This might lead to resources wasted on normal branches. 

In this paper, we try to tackle these challenges from three aspects: (1) In generating function invocation sequences, we explicitly consider data dependencies between functions to facilitate exploring richer states. We further prolong a function invocation sequence $\mathcal{S}_1$ by appending a new sequence $\mathcal{S}_2$, so that the appended sequence $\mathcal{S}_2$ can start fuzzing from states that are different from the initial state. (2) We incorporate a {branch distance}-based measure to evolve test cases iteratively towards a target branch.  (3) We engage a {branch search} algorithm to discover rare and vulnerable branches, and design an energy allocation mechanism to take care of exercising these crucial branches. We implement {IR-Fuzz} and extensively evaluate it over 12K real-world contracts. Empirical results show that: (i) {IR-Fuzz} achieves $28\%$ higher branch coverage than state-of-the-art fuzzing approaches, (ii) {IR-Fuzz} detects more vulnerabilities and increases the average accuracy of vulnerability detection by $7\%$ over current methods, and (iii) {IR-Fuzz} is fast, generating an average of $350$ test cases per second. Our implementation and dataset are released at \url{https://github.com/Messi-Q/IR-Fuzz}, hoping to facilitate future research. 
\end{abstract}


\section{Introduction}
\IEEEPARstart{S}{mart} contracts are programs executing on top of a blockchain system~\cite{zheng2018blockchain}. A smart contract encodes predefined contract terms into runnable code. Due to the immutable nature of blockchain, once a smart contract is deployed on the blockchain, its defined rules will be strictly followed during execution. Smart contracts make the automatic execution of contract terms possible, giving rise to a variety of  decentralized applications~\cite{liu2021combining,qian2019digital}. 

{Notably, not all blockchains support smart contracts. Ethereum, one of the most prominent blockchains enabling the execution of smart contracts, has attracted widespread attention worldwide.} So far, tens of millions of contracts have been deployed on Ethereum~\cite{wood2014ethereum}, enabling a broad spectrum of applications, including wallet~\cite{chen2019cryptoar}, crowdfunding~\cite{crowdfunding}, supply chain~\cite{mirabelli2020blockchain}, and cross-industry finance~\cite{industry}. Smart contracts from various fields now hold over one thousand billion dollars worth of virtual coins, and the number of contracts is still increasing rapidly~\cite{liu2021combining}. Smart contracts have long been appealing targets for attackers since they manipulate so many digital assets. 

Specifically, the source code of a Ethereum smart contract will be compiled into bytecode and executed on Ethereum Virtual Machine~\cite{rodler2021evmpatch}. Like traditional programs, smart contracts may contain vulnerabilities. Therefore, it is important to identify potential vulnerabilities in smart contracts, ideally before their deployments.  Malicious attackers may exploit the bugs in smart contracts to gain illegal profits. Recently, {there was an increasing number of security vulnerability incidents in smart contracts~\cite{qian2022smart,kushwaha2022systematic}}, leading to enormous financial losses. One infamous example was the reentrancy attack, \emph{i.e.,} attackers stole more than \$130 million worth of digital assets, exploiting the reentrancy vulnerability in the Cream.Finance contract~\cite{Inspex}. This case is not isolated, \emph{e.g.,} a delegatecall bug accidentally triggered resulted in freezing over \$280 million worth of Ether in the {Parity Multisig Wallet} contract~\cite{Parity}. Obviously, conducting security vetting on smart contracts to avoid exposing vulnerabilities to attackers is much coveted.

Fueled by the maturity of static analysis techniques such as formal verification~\cite{Bhargavan} and symbolic execution~\cite{oyente}, smart contract vulnerability detection has undergone considerable progress in the past few years. These methods, however, inherently suffer from high false positive rates since they do not actually execute each path. {Recent efforts resort to fuzzing techniques~\cite{nguyen2020sfuzz,zong2020fuzzguard,li2020v}, which have the merits of producing negligible false positives in discovering software vulnerabilities. This can be attributed to the fact that fuzzers usually generate test cases to exercise a branch, and report vulnerabilities only when they detect abnormal results during fuzzing.}

After scrutinizing existing released fuzzers for smart contracts, such as~\cite{li2020v,wustholz2020targeted,contractfuzzer,nguyen2020sfuzz,wustholz2020harvey,he2019learning,torres2021confuzzius}, we obtain the following observations. 
\textbf{(1)} Current fuzzers (\emph{e.g.,} {sFuzz}~\cite{nguyen2020sfuzz} and {Harvey}~\cite{wustholz2020harvey}) tend to  generate function invocation sequences randomly, overlooking the data dependencies (such as \emph{read} and \emph{write} dependencies) between functions. {More importantly}, a smart contract may transition through many different states during its lifecycle~\cite{wustholz2020harvey}. For example, every bet in a gambling contract will change the contract state. However, current methods generally conduct fuzzing starting from the initial state of the contract, which actually expends too much energy revolving around the initial state of the contract and is incapable of unearthing bugs triggered by complex states. \textbf{(2)} Most current approaches fail to take into account the distance between test cases and branch conditions in seed mutation, resulting in generating seeds that have low probabilities to enter a target branch. \textbf{(3)} Existing fuzzers (\emph{e.g.,} {ILF}~\cite{he2019learning} and {Confuzzius}~\cite{torres2021confuzzius}) often treat program branches equally. As a result, fuzzers might waste too many resources in fuzzing normal branches and are unable to dive deep into crucial branches that are rare or more likely to have bugs.

To tackle these challenges, we propose {IR-Fuzz}, a fully automatic \underline{Fuzz}ing framework equipped with \underline{I}nvocation ordering and impo\underline{R}tant branch revisiting, for detecting security vulnerabilities in Ethereum smart contracts. In particular, {IR-Fuzz} consists of three key components. 

\textbf{Sequence Generation.}\quad 
Usually, there are multiple functions within a contract, we introduce a function-invocation-sequence generation strategy, which consists of function invocation ordering and sequence prolongation. 
Specifically, we build a data flow analyzer to capture the data flow dependencies of global variables and then define a rule to compute the order priority of each function call, inferring the ordered function invocation sequence. Further, we introduce a prolongation technique to extend the sequence, enforcing the fuzzer to tap into  unprecedented states.

\textbf{Seed Optimization.}\quad 
We also present a seed optimization paradigm, which drives the fuzzer to generate branch-condition-aware test cases. In practice, we employ a {branch distance}-based measure to select test cases according to how far a test case is from satisfying the condition (\emph{e.g.}, $x$$=$$=$$10$) of a just-missed branch\footnote{A just-missed branch stands for the unexplored \emph{if}-branch or \emph{then}-branch of a conditional statement (such as \emph{if} and \emph{require}) or a recurrent statement (such as \emph{for} and \emph{while}).}. Intuitively, the test case has a higher probability to enter the just-missed branch as the distance decreases. In this way, {IR-Fuzz} iteratively evolves test cases to get increasingly closer to satisfying the branch conditions, which boosts its ability to find a high-quality test case and achieve a higher branch coverage. 

\textbf{Energy Allocation.}\quad 
Finally, we design an energy allocation mechanism that takes into account the significance of a branch. Technically, we first propose a branch search algorithm to pick out rare branches and branches that are likely to have vulnerabilities. Then, we formulate a customized energy schedule and develop two rules to guide fuzzing energy allocation. As such, IR-Fuzz can flexibly assign fuzzing resources to more important branches, which increases the overall  fuzzing efficiency by a large margin (4.9x faster than \emph{sFuzz}~\cite{nguyen2020sfuzz}) and further improves branch coverage. 

We implement {IR-Fuzz} and extensively evaluate this system over 12K real-world smart contracts. Experimental results show that: \textbf{(i)} {IR-Fuzz} achieves high average branch coverage by up to $90\%$, yielding a $28\%$ improvement compared with state-of-the-art fuzzing approaches. \textbf{(ii)} {IR-Fuzz} identifies more vulnerabilities and increases the average accuracy of vulnerability detection by $7\%$ over current methods. \textbf{(iii)} {IR-Fuzz} generates an average of $350$ test cases per second, in most cases orders-of-magnitude faster than conventional fuzzers.

Our \textbf{key contributions} can be summarized as follows:
\begin{itemize}[topsep=1.5pt]
\item We design and implement a novel framework {{IR-Fuzz}} for smart contract fuzzing, which consists of three key components, {i.e.}, \emph{function invocation sequence prolongation}, \emph{branch-distance-driven seed optimization}, and \emph{branch-importance-aware energy allocation}. 
\item Within the framework, we present a sequence generation strategy to infer high-quality function invocation sequences, steering fuzzing to explore unprecedented states. Further, we introduce a seed optimization paradigm that incorporates a {branch distance}-based measure to select and evolve test cases towards new branches. Finally, we develop a branch search algorithm to discover rare and vulnerable branches, and propose an energy allocation mechanism to concentrate on these critical branches.
\item We evaluate {IR-Fuzz} over large-scale real world smart contracts, and empirical results show that the proposed techniques are indeed useful in achieving high branch coverages. IR-Fuzz surpasses state-of-the-art fuzzers and overall provides interesting insights. As a side contribution, we construct a large benchmark dataset for evaluating smart contract fuzzing approaches. Our implementation and dataset are released, hoping to inspire others. 
\end{itemize}

\section{Related Work}

\subsection{Smart Contract Vulnerability Detection}
Since blockchain endows smart contracts with unalterable nature, there is no way to patch the vulnerabilities of a smart contract without forking the blockchain (almost an impossible mission), regardless of how much money the contract holds or how popular it is~\cite{Sok,zhuangsmart,qian2020towards,liu2021combining}. Therefore, it is critical to conduct security vetting for smart contracts, especially before their deployments.  Early works for smart contract vulnerability detection employ formal verification techniques. For example,~\cite{Bhargavan} introduces a framework to compile smart contracts to EVM bytecode and then put them into an existing verification system.~\cite{Hirai} proposes a formal model and verifies smart contracts using the Isabelle/HOL tool. Further,~\cite{Grishchenko} and~\cite{Hildenbrandt} define the formal semantics of contracts using the F* framework and the $\mathbb{K}$ framework, respectively. Although these frameworks provide strong formal verification guarantees, they are still semi-automated and yield high false positives. Another stream of works rely on symbolic execution methods, such as Oyente~\cite{oyente}, Slither~\cite{feist2019slither}, and Securify~\cite{securify}. Oyente is one of the pioneering works to perform symbolic execution on smart contracts, which checks bugs based on expert-defined rules.~\cite{feist2019slither} converts smart contracts into intermediate representations and conducts static analysis to detect vulnerabilities. Whereas symbolic execution is a powerful technique for discovering bugs, it still suffers from the inherent problem of symbolic execution path explosion. 

Recent efforts resort to using fuzzing techniques for smart contract vulnerability detection. ContractFuzzer~\cite{contractfuzzer} is the first to apply fuzzing techniques to smart contracts and identifies vulnerabilities by monitoring runtime behaviors during fuzzing. ReGuard~\cite{reguard} and Harvey~\cite{wustholz2020harvey} are dedicated to generating a number of test cases that cover as many paths as possible to trigger a vulnerability. ILF~\cite{he2019learning} and sFuzz~\cite{nguyen2020sfuzz} aim to design a feedback-based seed mutation strategy. Despite the practicality of fuzzing techniques, existing fuzzers still have difficulties in achieving high coverage and fuzzing efficiency. Instead, our work alleviates the issues by carefully designing a sequence generation strategy, a seed optimization paradigm, and an energy allocation mechanism.

\subsection{Greybox Fuzzing} 
Fuzzing techniques have been proven as an effective way to discover software vulnerabilities. According to how much information is available about the  program under test~\cite{liang2018fuzzing}, fuzzing techniques can be cast into three categories: \textit{whitebox}, \textit{blackbox}, and \textit{greybox}~\cite{zheng2019firm,bohme2017directed,xu2019fuzzing}. Put succinctly, blackbox testing conducts fuzzing without knowing any internal structure of the target program. In contrast, whitebox testing performs fuzzing while having full knowledge about the internal architecture of the target program. Greybox fuzzing stands in the middle of blackbox fuzzing and whitebox fuzzing, where we have partial knowledge of the internal structure of the target program. Particularly, \emph{greybox} fuzzing can be further divided into two groups.

One spectrum of works~\cite{chen2018angora,gan2018collafl} aim at covering as many paths or branches as possible, expecting to reveal a bug in the program, namely \emph{coverage-guided greybox fuzzing}. AFL, one of the most well-known fuzzers, employs the lightweight instrumentation technique and genetic algorithm to improve coverage~\cite{AFL}. Some other researchers~\cite{bohme2017coverage,nguyen2020sfuzz} increase code coverage by smartly selecting and mutating test cases. Typically, these methods improve coverage by generating as many new test cases as possible to traverse previously uncovered program paths. Another spectrum of works~\cite{bohme2017directed,chen2018hawkeye,chen2018angora} are designed to direct greybox fuzzing towards a set of specific target locations, termed \emph{targeted greybox fuzzing}. There is a number of greybox fuzzers that focus on specific program locations, \emph{e.g.,} low-frequency or uncovered branches. For example,~\cite{wustholz2020targeted} utilizes a power schedule to collect feedback information and steer fuzzing towards target locations. AFLGo~\cite{bohme2017directed} calculates the distance between entry points and buggy code in the control flow graph, guiding seed mutation to cover the target locations. Overall, targeted greybox fuzzing generates test cases to reach certain target locations, attempting to further trigger a bug.

\section{Motivating Example}
\begin{figure}
\centering
\includegraphics[width=8.7cm]{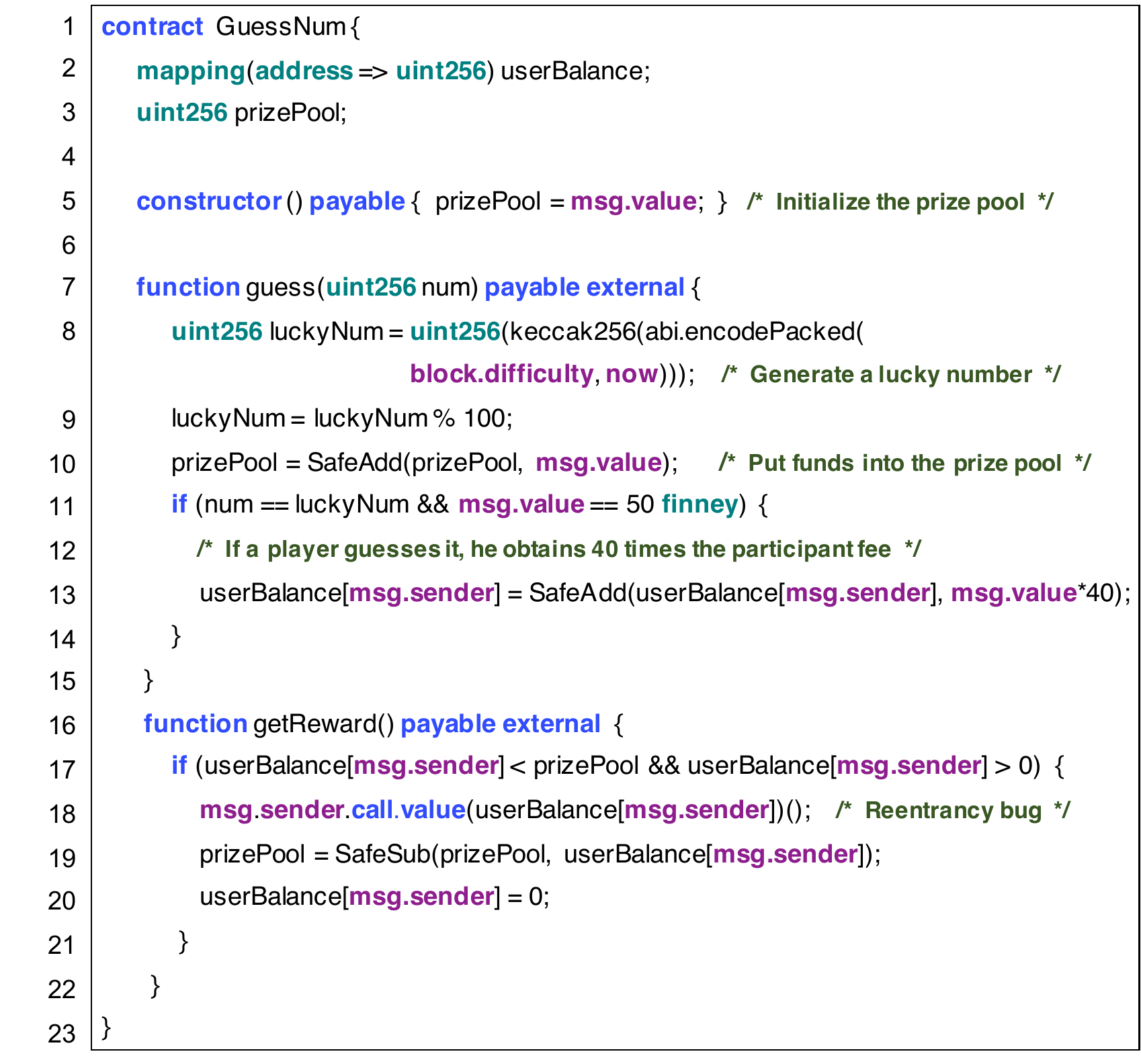}
\caption{A real-world smart contract written in Solidity.} 
\vspace{-0.8em}
\label{fig1}
\end{figure}

As a motivating example, we present a real world smart contract \emph{GuessNum}, which implements a gambling game on Ethereum~\cite{wood2014ethereum}. Fig.~\ref{fig1} shows the simplified code\footnote{ Address: \href{https://ropsten.etherscan.io/address/0xd5e94f6350c7911015dd120b0b006420b6e85a58\#code}{0xd5e94f6350c7911015dd120b0b006420b6e85a58} } of \emph{GuessNum}, which is written in Solidity. The contract realizes a {guess number} game that users play by submitting their guesses along with participation fees. The fee is fixed to 50 finney each time and is poured into the prize pool, \emph{i.e.,} \emph{prizePool}. Function \emph{constructor()} runs only once when the contract is created, and it puts the funds of the contract's owner into the prize pool. A player who wants to submit a \emph{guess} can invoke function \emph{guess()}, which compares the received \emph{guess} number with the randomly generated lucky number, \emph{i.e.,} \emph{luckyNum} (line 11). If \emph{guess} number exactly matches the \emph{luckyNum}, the player will obtain 40 times the participation fee in return (line 13). Players can get rewards by calling function \emph{getReward()}.

\textbf{Vulnerability.}\quad This \emph{GuessNum} contract, unfortunately, suffers from a classical reentrancy vulnerability. From line 18 of Fig.~\ref{fig1}, we observe that function \emph{getReward()} invokes \emph{call.value} to transfer money to the user. However, due to the default settings of smart contracts, the \emph{transfer} operation will automatically trigger the \emph{fallback} function of the recipient contract. Therefore, an attacker may set a malicious second-time invocation to \emph{getReward()} in his \emph{fallback} function for stealing extra money. Since \emph{getReward()} waits for the first-time \emph{transfer} to finish, the balance of the attacker is not reduced yet (\emph{i.e.,} the user balance reduction operation at line 19 is behind \emph{call.value} and is not executed yet). Function \emph{getReward()} thus may wrongly believe that the attacker still has enough balance and transfers money to the attacker again.

\textbf{Limitation of Existing Fuzzers.}\quad 
Interestingly, this simple smart contract reveals three key challenges for most existing fuzzers to expose  vulnerabilities. 
\textbf{(1)} The order of function invocations is critical. We observe that if the conditions at line 11 are not satisfied (\emph{namely} the then-branch at line 13 is not reached), then the second condition at line 17 will never be met either (because \textit{userbalance[msg.sender]} is equal to 0). As such, the fuzzer will be unable to reach the then-branch at lines 18--20. 
\textbf{(2)} Generating a test case to satisfy the second condition at line 11 (\emph{msg.value == 50 finney}) is difficult. More specifically, the variable \emph{msg.value} has a size of 32 bytes. Thus, when we generate a random value for \emph{msg.value} in fuzzing, we have only $\frac{1}{2^{256}}$ probability to obtain the value 50 to meet the condition. Indeed, existing fuzzers like AFL~\cite{AFL} are shown to have difficulties in electing a test case to enter the then-branch at line 13. \textbf{(3)} Vulnerable branches that may contain vulnerabilities only take a small fraction of the program. For example, lines 18--20 are vulnerable code which only exists in one branch. Current fuzzers~\cite{he2019learning,torres2021confuzzius} tend to treat each branch equally, which may fail to discover the vulnerability due to insufficient fuzzing resource allocation.

\begin{figure*}
    \centering
    \includegraphics[width=18.2cm]{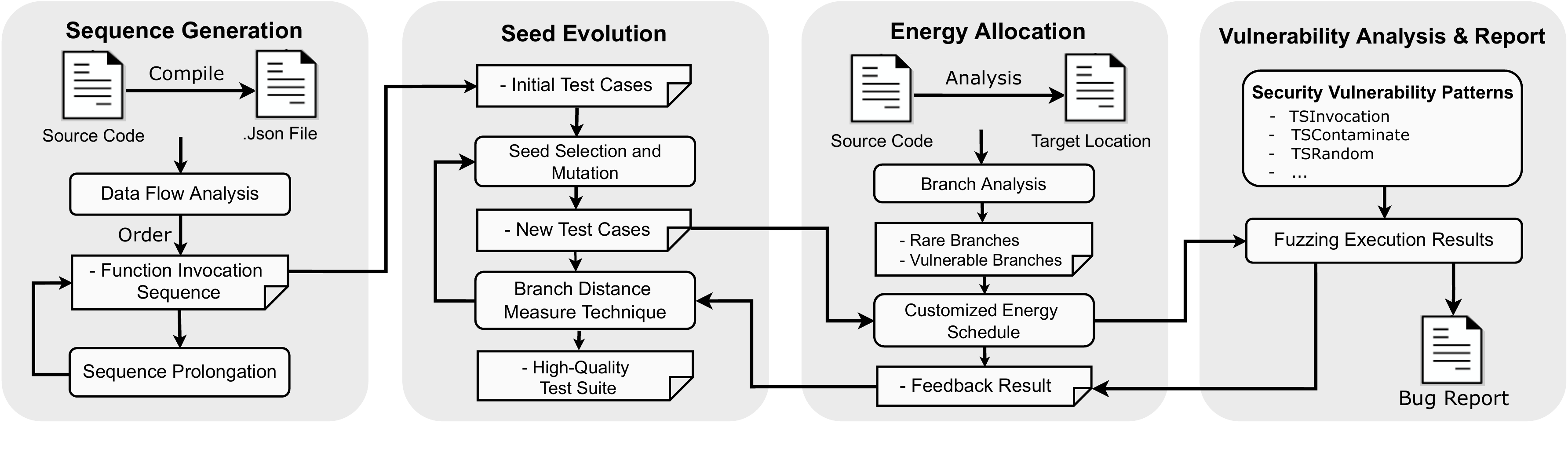}
    \caption{
       A high-level overview of {IR-Fuzz}. {IR-Fuzz} has four main components, including (1) Sequence Generation, (2) Seed Optimization, (3) Energy Allocation, and (4) Vulnerability Analysis and Report. 
    }
    \vspace{-0.8em}
     \label{fig:overview}
\end{figure*}

\textbf{Fuzzing Policy.}\quad 
We embrace three key designs in {IR-Fuzz} to tackle the  challenges. 
\textbf{(1)} {IR-Fuzz} leverages the variable \emph{read} and \emph{write} dependencies between functions to generate the ordered function invocation sequence. It further extends the ordered sequence with another ordered sequence to explore more complex states. Specifically, this contract (Fig.~1) has two global variables, {i.e.,} \emph{userBalance} and \emph{prizePool}, which both appear in functions \emph{guess()} and \emph{getReward()}. By analyzing \emph{read} and \emph{write} dependencies of the global variable \emph{userBalance}, IR-Fuzz recognizes that function \emph{getReward()} depends on function \emph{guess()}\emph, awaring that \emph{guess()} should be called before \emph{getReward()}. Consequently, IR-Fuzz generates the function invocation sequence as: \emph{guess()}$\rightarrow$\emph{getReward()}. 
\textbf{(2)} {IR-Fuzz} adopts a {branch distance}-based schema to select test cases according to how far a test case is from satisfying the condition of a just-missed branch. For example, the distance of reaching the just-missed branch (\emph{i.e.,} then-branch at line 13) is calculated as $\lvert$\emph{msg.value} - 50$\rvert$ since the branch condition at line 11 is \emph{msg.value == 50}. Intuitively, the test case has a higher probability to enter the just-missed branch as the distance decreases. With the guidance of distance measure, {IR-Fuzz} iteratively evolves test cases to get increasingly closer to satisfying the branch condition at line 11. 
\textbf{(3)} {IR-Fuzz} engages a branch search algorithm to pick out vulnerable (\emph{e.g.,} then-branch at lines 18--20) and rare branches, and then formulates an energy schedule to expend more fuzzing resources on these important branches. In our experiments, after only 26s, {IR-Fuzz} generates a test case to reach the then-branch at line 18 and exposes the reentrancy vulnerability.

\section{Method}
\label{method}
\textbf{Overview.}\quad 
The overall architecture of IR-Fuzz is outlined in Fig.~\ref{fig:overview}. Generally, IR-Fuzz consists of four key components:
\begin{itemize}[topsep=1pt, leftmargin=\dimexpr\labelwidth + 2\labelsep\relax]
\item \emph{Sequence Generation}: Given that a contract might contain multiple functions, to explore their possible function invocation sequences, {IR-Fuzz} first analyzes the data flow dependencies between functions. Then, it defines a rule to compute the order priority of each function, and generates a function invocation sequence that successively calls the ordered functions. Further, IR-Fuzz adopts a prolongation technique to extend the sequence, driving the fuzzer to dive into deeper states.
\item \emph{Seed Evolution}: To guide seed mutation so that the generated cases could reach a target branch, IR-Fuzz utilizes  a {branch distance}-based measure to select and evolve test cases iteratively according to how far a test case is from satisfying the condition of the target branch. 
\item \emph{Energy Allocation}: To further take care of the rare branches and branches that are likely to have vulnerabilities, {IR-Fuzz} introduces a branch search algorithm to analyze exercised branches and picks out those important branches. Then, {IR-Fuzz} formulates a customized energy schedule and utilizes two rules to flexibly guide fuzzing energy allocation towards these critical branches.
\item \emph{Vulnerability Analysis and Report}: {IR-Fuzz} analyzes the generated logs and refers to vulnerability-specific patterns to discover  vulnerabilities. Bug reports are generated for further manual inspections.
\end{itemize} 
In what follows, we will elaborate on the details of these components one by one.

\subsection{Sequence Generation}
\label{order_sequence}

Presented with the multiple functions of a smart contract, existing methods tend to generate a function invocation sequence by randomly picking a function each time. Scrutinizing 12K real-world smart contracts, we empirically observe that the state of a smart contract is often captured by the state of its global variables, and different functions do share and operate differently on the global variables. Some functions perform `read' operations on the variables while some other functions may perform `write' operations on the variables. Generating function invocation sequence  randomly ignores such connections between functions. The functions that perform `write' operations could change the state of the contract while the functions that perform only `read' operations are unable to change the state. Therefore, we assign higher order priority to functions that perform `write' operations so that we may explore a broader contract state space and cover more branches. It is worth mentioning that initializing a variable will not affect the method.

\textbf{Sequence Ordering.}\quad 
Motivated by this, we propose to model the global variable \emph{read} and \emph{write} dependencies between functions. Specifically, we first characterize the source code of a smart contract into an abstract syntax tree~\cite{zhang2019novel}, from which we extract variable access operations such as assignments and comparisons. Then, we leverage a data flow analyzer to capture the {read and write} dependencies of global variables between functions. Afterwards, we calculate the order priority (OP) of each function and sort them according to their OPs. Finally, we generate the ordered function invocation sequence, \emph{i.e.}, the invocation sequence that calls the sorted functions successively. 

Formally, we denote the set of functions in a smart contract as $\mathcal{F}=\{F_1, F_2, ..., F_N\}$, and global variables that appear in function $F_i$  as $\mathcal{V}=\{v_{i1}, v_{i2}, ..., v_{iM}\}$, where $M$ is the number of global variables appear in $F_i$ and $v_{ik}$ represents the $k$-th global variable in $F_i$. Specifically, each variable  $v_{ik}$  has a unique identifier (\emph{i.e.}, variable name), which is denoted as $v^{ID}_{ik}$. To further indicate the operation that  $F_i$ exerted on $v_{ik}$, we use $v^{op}_{ik} = 1$ to represent that $F_i$ performs \emph{read} operation on  $v_{ik}$, and $v^{op}_{ik} = 0$ to denote that $F_i$ conducts \emph{write} operation. Now, we define the order priority of functions as below. 

\vspace{3px}
\textbf{Rule 1: Read \& Write Dependency.}\quad \emph{When a global variable appears in two different functions, the function that executes write operation on the variable should be called earlier than the function that executes read operation. Put differently, given global variables $v_{ik}$ in $F_i$ and $v_{jn}$ in $F_j$ (where $v^{ID}_{ik}$ = $v^{ID}_{jn}$), we suggest that $F_i$ should have a higher order priority ($\textbf{OP}_{F_i} > \textbf{OP}_{F_j}$) when $v^{op}_{ik} = 0$ and $v^{op}_{jn} = 1$. }
\vspace{3px}

Guided by this, we may compute the order priority of each function by converting this rule into the following formula, which sums up the analysis on \emph{read \& write} dependencies (\textbf{Rule 1}) of all global variables in different functions.
\begin{equation}
\label{eq1}
\small
    \begin{split}
        \textbf{OP}_i = \sum_{k = 1}^{M_i} & \sum_{p=1}^{M_j} 
        v^{op}_{jp} 
        (1 -  v^{op}_{ik})
       \cdot cmp( v_{ik},  v_{jp}) \\
        &1 \leq i,j \leq N \ \ \ \  \&\& \ \ \ \  i \neq j
    \end{split}
\end{equation}
where $N$ is the number of functions in the contract. $M_i$ and $M_j$ denotes the number of the appearance of global variables in $F_i$ and $F_j$, respectively. Notably, $cmp(v_{ik}, v_{jp})$ compares the identifier of global variables in two different functions, which is given by:
\begin{equation}
\small
    cmp(v_{ik}, v_{jp}) = \begin{cases}
        1, & v^{ID}_{ik} = v^{ID}_{jp}    \\
        0, & v^{ID}_{ik} \neq v^{ID}_{jp}  
    \end{cases}
\end{equation}
where $v_{ik}$ denotes $k$-th variable of $F_i$ and $v_{jp}$ represents $p$-th variable of $F_j$. 

Mathematically, in Eq.(\ref{eq1}), a function accumulates one order priority score only if it writes on a global variable and the variable is read by another function. In this context, the functions that conduct write operations on more global variables get higher priority and are put in the front of the function invocation sequence. This drives the fuzzer to exercise more on the functions that could change the states and boost the fuzzing by encouraging it to encounter more states and reach more branches. Interestingly, our experimental results show that branch coverage is significantly improved with such sequence ordering (see \S\ref{ablation_study}). 
Here, we take the contract of Fig.~\ref{fig1} as an example to illustrate the order priority calculation of each function. From Fig.~\ref{fig1}, we can observe that this contract has two global variables, \textit{i.e.,} \textit{userBalance} and \textit{prizePool}. Function \textit{guess()} performs a \textit{read} operation and a \textit{write} operation on the two variables, respectively. We represent the  variables that appear in function \textit{guess()} as $\mathcal{V}_{\textit{guess}}$ $=$ $\{v_{\textit{guess}1}, v_{\textit{guess}2}, v_{\textit{guess}3}, v_{\textit{guess}4} \}$, where $v_{\textit{guess}1}$ $=$ $v_{\textit{guess}2}$ $=$ \textit{prizePool}, $v_{\textit{guess}1}^{op}$ $=$ $1$, $v_{\textit{guess}2}^{op}$ $=$ $0$, and $v_{\textit{guess}3}$ $=$ $v_{\textit{guess}4}$ $=$ \textit{userBalance}, $v_{guess3}^{op} = 1$, $v_{guess4}^{op} = 0$. Meanwhile, ${\textit{cmp}}$\textit{(userBalance}, \textit{prizePool)} = 0, while ${\textit{cmp}}$\textit{(userBalance}, \textit{userBalance)} =  ${cmp}$\textit{(prizePool}, \textit{prizePool)} = 1. Similarly, the  global variables appear in \textit{getReward()} are denoted as $\mathcal{V}_{\textit{getReward}}$ = $\{v_{\textit{getReward}1}, v_{\textit{getReward}2}, ..., v_{\textit{getReward}8} \}$, where $v_{\textit{getReward}1}^{op}$ = ... = $v_{\textit{getReward}6}^{op}  = 1$, and $v_{\textit{getReward}7}^{op} = v_{\textit{getReward}8}^{op}  = 0$. As such, according to Eq.(\ref{eq1}), we calculate the order priority of function \textit{guess()} as below.
\begin{equation}
  \small
    \begin{split}
        \textbf{OP}_{\textit{guess}}  =&  v_{\textit{getReward}1}^{op}(1 - v_{\textit{guess}4}^{op})  
        +  v_{\textit{getReward}2}^{op}(1 - v_{\textit{guess}2}^{op}) \\
        +&  v_{\textit{getReward}3}^{op}(1 - v_{\textit{guess}4}^{op}) 
        +  v_{\textit{getReward}4}^{op}(1 - v_{\textit{guess}4}^{op}) \\
        +&  v_{\textit{getReward}5}^{op}(1 - v_{\textit{guess}2}^{op}) 
        +  v_{\textit{getReward}6}^{op}(1 - v_{\textit{guess}4}^{op}) \\
        =  & 6                                                      
    \end{split}
\end{equation}
For function \textit{getReward()}, we calculate its order priority as:
\begin{equation}
\small
    \begin{split}
        \textbf{OP}_{\textit{getReward}} = & v_{\textit{guess}1}^{op}(1 - v_{\textit{getReward}7}^{op})   \\
        +&  v_{\textit{guess}3}^{op}(1 - v_{\textit{getReward}8}^{op}) \\
        =& 2
    \end{split}
\end{equation}
According to the order priority calculation, we can see that the order priority of calling function \textit{guess()} is greater than that of function \textit{getReward()}. Therefore, {IR-Fuzz} finally generates the function invocation sequence as: \textit{guess()} $\rightarrow$ \textit{getReward()}.

\begin{figure}
    \centering
    \includegraphics[width=8.7cm]{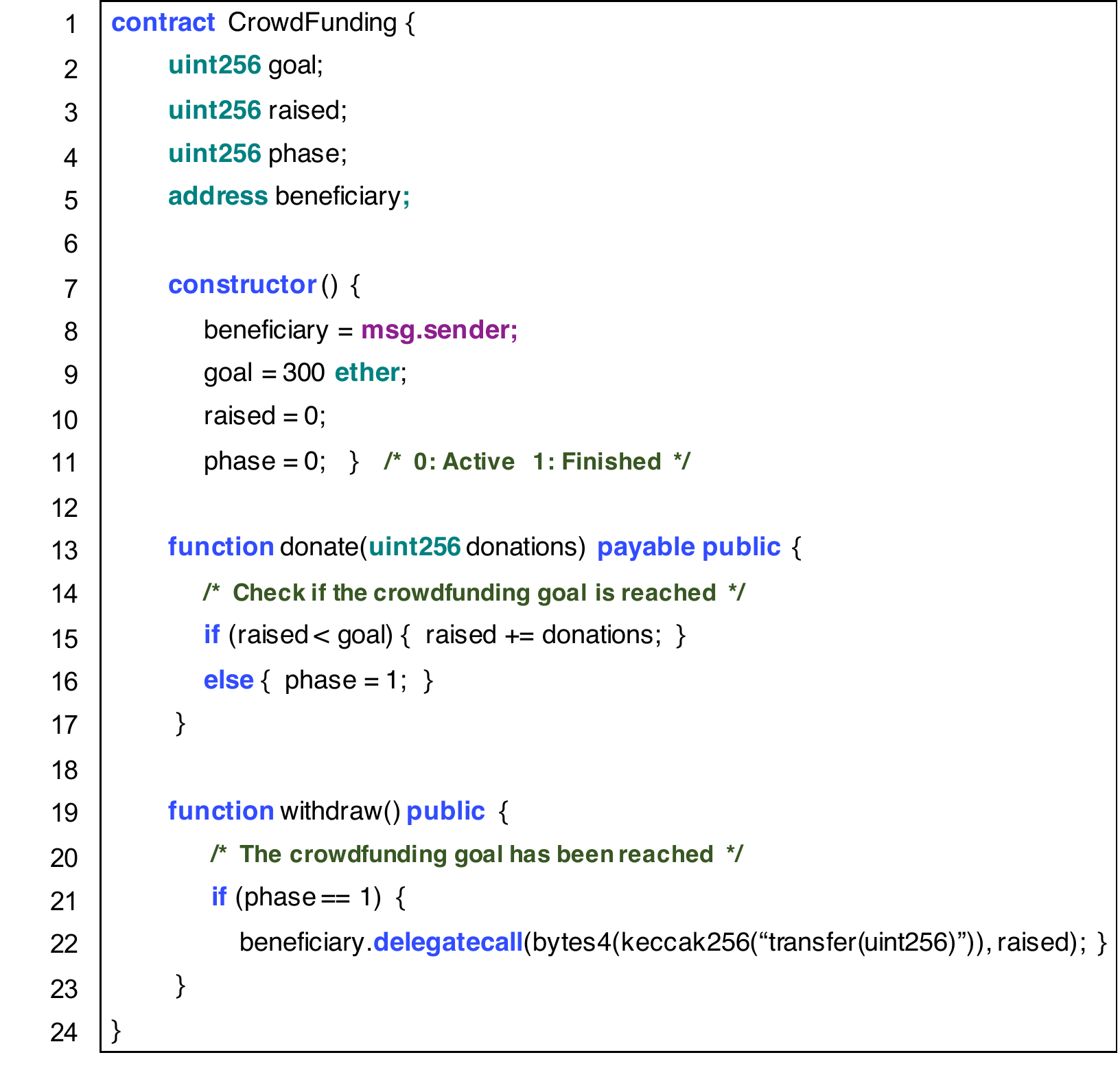}
    \caption{An example contract for illustrating how the sequence prolongation technique is used in IR-Fuzz.}
    \label{fig_crowdfund}
    \vspace{-0.8em}
\end{figure}

\textbf{Sequence Prolongation.}\quad 
Another important insight is that a smart contract might go through many different states during its lifecycle. However, current methods typically conduct fuzzing starting from the initial state of the contract, which expends too much energy revolving around the initial state and is usually incapable of unearthing bugs triggered by other states. These facts inspire us to explore richer starting states via sequence prolongation. Particularly, we first exercise the ordered function invocation sequence $\mathcal{S}$ using various test cases, which result in different states of the contract. We then engage in the invocation sequence $\mathcal{S}$ again but execute $\mathcal{S}$ starting from these different states, \emph{i.e.}, appending a new sequence $\mathcal{S}$ after $\mathcal{S}$.  

For instance, presented with the crowdfunding contract in Fig.~\ref{fig_crowdfund}, we observe that covering the if-branch at line 21 for traditional fuzzers is difficult, which requires at least \emph{twice} invocations of function \emph{donate()}. Particularly, this contract implements a simple crowdfunding project that allows users to donate money by calling \emph{donate()}. The \emph{constructor()} sets the goal of crowdfunding as 300 Ether (line 9), and the raised money is  initialized to 0 (\emph{i.e.,}  \emph{raised = 0} at line 10). The status of the crowdfunding process is initialized to 0 (\emph{i.e.,}  \emph{phase} = 0 at line 11), which represents \emph{unaccomplished}. According to \textbf{Rule 1}, we realize   that the invocation to function \emph{donate()} should have a higher {order priority} than that of function \emph{withdraw()}. As a result, the function invocation sequence is generated as: \emph{donate()}$\rightarrow$ \emph{withdraw()}. However, such a sequence fails to satisfy the branch condition at line 21 because call function \emph{donate()} {once} cannot enter the else-branch at line 16 to set \emph{phase} = 1. To reach the else-branch at line 16, function \emph{donate()} needs to be called at least \emph{twice}. Namely, the required amount of money for the crowdfunding is accomplished at the first call (\emph{i.e.,} \emph{raised} $\geq$ \emph{goal}), and the second call enters the else-branch at line 16. Therefore, to explore deeper states, we propose to prolong the function invocation sequence. 

In contrast to most fuzzing methods that conventionally call each function only once in the invocation sequence, we further extend the ordered function invocation sequence $\mathcal{S}$ by appending the same invocation sequence to $\mathcal{S}$, \emph{namely} $\mathcal{S}\rightarrow\mathcal{S}$,  such that the second sequence starts its execution from a different state rather than the initial state. To explore different starting states for the second sequence, the first sequence is presented with different sets of parameters, which lead to different inner statuses. Technically, we first try to sufficiently  exercise sequence $\mathcal{S}$ with various different input parameters. A different set of input parameters to $\mathcal{S}$ translates to a variant of $\mathcal{S}$, denoted as $\mathcal{S}_j$. Then, we concatenate two sequences $\mathcal{S}_i$ and $\mathcal{S}_j$ that have different input parameters and exercise the new concatenated sequence. The following rule formulates the sequence pair selection to generate a new prolonged sequence, which constrains that the input parameters of $\mathcal{S}_i$ and $\mathcal{S}_j$ should be quite different.

\begin{algorithm}[t]
    \small    
    \setstretch{0.88} 
    \caption{ \textsc{Seed Iterative Optimization} }
    \label{alg1}
    
    $currentTestSuite \leftarrow \varnothing$\;
    $currentTestCase \leftarrow initialTestCase()$\;
    
    \While{$\lnot Terminated()$}{
        Let $B_{testCase}$ be covered branches by $currentTestCase$\;
        Let $B_{miss}$ be just-missed branches in $currentTestSuite$\;
        \For{$b_r \in B_{testCase}$}{
            \If{$b_r$ is new branch}{
                $currentTestSuite$.ADD($currentTestCase$)\;
            }
        }
        \For{$b_r \in B_{miss}$}{ 
            \For{$seed \in currentTestSuite$ \&\& \hspace*{0.12in}$seed \neq currentTestCase$}{
                $dist\_1 \leftarrow distance(currentTestCase, b_r)$\;
                $dist\_2 \leftarrow distance(seed, b_r)$\;
                \If{$dist\_1 < dist\_2$}{
                    $currentTestSuite$.REMOVE($seed$)\;
                    $currentTestSuite$.ADD($currentTestCase$)\;
                }
            }
        }
       $energy \leftarrow 0$\;
       $MutationEnergy \leftarrow AssignMutationEnergy()$\;
        \While{$energy < MutationEnergy$}{
            $testCase \leftarrow SelectInput(currentTestSuite)$\;
            $newCase \leftarrow Mutation(testCase)$\;
            \If{$\lnot RepeatCheck(currentTestSuite, newCase)$ \&\& \hspace*{0.05in}$\lnot ValidityCheck(newCase)$}{
                $currentTestCase$.ADD($newCase$)\;
                $log \leftarrow FuzzInput(newCase)$\;
                $energy \leftarrow UpdateEnergy(log, energy)$\;
            }
        }
    }
     \Return{ $currentTestSuite$: A set of high-quality test cases}
\end{algorithm}

\vspace{3px} 
\textbf{Rule 2: Sequence Pair Selection.}\quad \emph{(1) When the number of required function input parameters in the sequence is no less than 2,  we select the sequence pairs \emph{iff} at least \emph{one} input paramter is different between the two sequences.  For example, given three sequences $\mathcal{S}_1$: $F_1(x_1) \rightarrow F_2()$, $\mathcal{S}_2$: $F_1(x_2) \rightarrow F_2()$, and $\mathcal{S}_3$: $F_1(x_2) \rightarrow F_2()$,  pairs $\mathbb{P}_1$($\mathcal{S}_1$, $\mathcal{S}_2$) and $\mathbb{P}_2$($\mathcal{S}_1$, $\mathcal{S}_3$) are selected. In contrast, pair $\mathbb{P}_3$($\mathcal{S}_2$, $\mathcal{S}_3$) is not selected since $\mathcal{S}_2$ and $\mathcal{S}_3$ have the same parameters.  (2) When the number of required function input parameters is larger than 2, we select the sequence pairs \emph{iff} at least \emph{two} input parameters are different between the two  sequences. For example, given three sequences $\mathcal{S}_1$: $F_1(x_1, y_1) \rightarrow F_2(z_1)$, $\mathcal{S}_2$: $F_1(x_1, y_2) \rightarrow F_2(z_1)$, and $\mathcal{S}_3$: $F_1(x_2, y_2) \rightarrow F_2(z_2)$, pairs $\mathbb{P}_1$($\mathcal{S}_1$, $\mathcal{S}_3$) and $\mathbb{P}_2$($\mathcal{S}_2$, $\mathcal{S}_3$) are selected.} 
\vspace{3px}

Upon each sequence pair selection, we obtain a prolonged function invocation sequence. For the crowdfunding contract demonstrated in Fig.~\ref{fig_crowdfund}, {IR-Fuzz} first generates two sequences, \emph{e.g.,} $\mathcal{S}_1$: \emph{donate(300)} $\rightarrow$ \emph{withdraw()} and $\mathcal{S}_2$: \emph{donate(200)} $\rightarrow$ \emph{withdraw()}. According to \textbf{Rule 2}, {IR-Fuzz} combines $\mathcal{S}_1$ and $\mathcal{S}_2$ as a new prolonged sequence $\mathcal{S}$: \emph{donate(300)} $\rightarrow$ \emph{withdraw()}$\rightarrow$ \emph{donate(200)} $\rightarrow$ \emph{withdraw()}. Since the goal of crowdfunding is 300, the first call to \emph{donate(300)} in $\mathcal{S}$ satisfies the condition of the if-branch at line 15, and the second call to \emph{donate(200)} reaches the else-branch at line 16 (\emph{i.e.,} set \emph{phase} = 1). The new sequence thus can reach the then-branch at line 22 on the second call to \emph{withdraw()} and expose the potential \emph{Ether frozen} vulnerability. Promisingly, with the prolongation technique, {IR-Fuzz} greatly expands the scope of explored states and branches. Note that the sequence prolongation technique yields little impact on the overhead of the fuzzer.

\subsection{Seed Evolution} 
\label{seed_optimization}
To fuzz a function invocation sequence, the most intuitive and direct way is to generate test cases randomly. Despite its simplicity, this strategy is not favorable for reaching unexplored conditional branches due to its random nature. {IR-Fuzz}, instead, incorporates a seed evolution paradigm to refine test cases iteratively. The seed evolution framework is summarized in Algorithm~\ref{alg1}. First, {IR-Fuzz} initializes an empty test suite and a set of test cases (lines 1-2). Then, the first loop from lines 6 to 8 performs seed selection. Whenever a test case covers a new branch (\emph{i.e.,} any branch not covered by test cases in the test suite), it is added to the suite. Next, the loop from lines 9 to 15 evolves test cases iteratively. Particularly, we propose a {branch distance}-based measure to select those test cases which are closer to satisfying the conditions of  new branches. Thereafter, the loop from lines 18 to 24 executes seed mutation, in which function \emph{Mutation()} generates the mutated test cases based on the test cases selected from the suite (lines 19-20). Then, we adopt seed verification strategies to guarantee the validity of mutated test cases (line 21). This mutation process continues until a mutation energy upper-bound is reached (line 18). Finally, a new test suite that contains a set of high-quality test cases is shaped. In what follows, we present the technical details of seed selection and seed mutation, respectively.

\textbf{Seed Selection.}\quad In {IR-Fuzz}, we first try a classical seed selection strategy. That is, {IR-Fuzz} monitors the execution of test cases and records the branches that each test case traverses. A test case is added into the test suite as long as it covers a new branch, \emph{i.e.,} a branch which is not covered by any test case in the suite. Empirically, our experimental results show that this strategy could reveal a number of branches. However, it is still quite inefficient in reaching those complex branches with strict conditions. For example, the probability of satisfying the second condition (\emph{msg.value} == 50 finney) at line 11 of Fig.~\ref{fig1} is $\frac{1}{2^{256}}$, which is extremely low. To meet such strict branch conditions, we design a novel seed selection strategy. Inspired by~\cite{nguyen2020sfuzz}, we adopt a distance function $dist(T, b_r)$ to compute a branch distance indicating how far a test case is from covering a just-missed branch (\emph{i.e.,} uncovered then-branch). More specifically, let $b_r$ be a just-missed branch, which is not covered by any test case $T$. We suppose that $b_r$ is a branch of condition $C$. Note that $C$ can be either $x$$==$$k$, $x != k$, $x\leq k$, $x<k$, $x \geq k$ or $x>k$, where $x$ and $k$ are variables or constants. The function $dist(T, b_r)$ is given by: 
\begin{equation}
\small
    dist(T, b_r) = \begin{cases}
        |x - k|, & if\; C\; is\; x == k                \\
        1,           & if\; C\; is\; x\ != k                    \\
        \max(x - k, 0)   & if\; C\; is\; x \leq k\; or\; x < k   \\
        \max(k - x, 0)  & if\; C\; is\; x \geq k\; or\; x > k 
    \end{cases}
    \label{eq:distance}
\end{equation}
where $x$ and $k$ are extracted from the stack information recorded by IR-Fuzz. Intuitively, $dist(T, b_r)$ is defined such that the closer a test case $T$ is from satisfying the condition of branch $b_r$, the smaller the distance is. For example, a test case with \emph{msg.value} = 100 is closer to satisfying the condition \emph{msg.value} == 50 than a test case with \emph{msg.value} = 10,000. For each just-missed branch, {IR-Fuzz} selects a test case has the smallest $dist(T, b_r)$. With the feedback of the branch distance measurement, IR-Fuzz can quickly approach complex branches guarded by strict conditions, improving the overall branch coverage. Note that all selected test cases are added to the test suite and transferred to the seed mutation phase for generating new test cases.

\textbf{Seed Mutation.}\quad 
Seed mutation plays an important role in enriching the test cases. In {IR-Fuzz}, we refer to several mutation strategies from AFL and introduce \emph{new} ones tailored for smart contracts. Particularly, we preferentially mutate those test cases with smaller branch distances.

Given a test case encoded in the form of a bit vector, sFuzz~\cite{nguyen2020sfuzz} engages a set of mutation operators to generate new test cases, such as bit flipping, interest value insertion, and key-value insertion. IR-Fuzz additionally ensures the generated test cases are valid by advocating two principles. 
\textbf{(1)} IR-Fuzz checks the validity and integrity of the mutated test cases by using a bit verification approach, which sets random bits of a given seed to random values while keeping other bits of the seed unchanged. Thereafter, IR-Fuzz saves a new test case as a new seed based on whether the new test case detects a new branch. Moreover, IR-Fuzz will discard invalid test cases which lead to fuzzing crashes or bring much overhead to the fuzzer.
\textbf{(2)} IR-Fuzz removes duplicate test cases by comparing the mutated cases with the test cases in the test suite. 

In practice, we also apply multiple heuristics to save the mutation energy of IR-Fuzz. For example, any test case in the test suite that does not discover a new branch after a round of mutation will be assigned with low priority in the mutation. {IR-Fuzz} updates \emph{energy} according to the generated logs during fuzzing (lines 23-24 in Algorithm~\ref{alg1}). 
The process of seed mutation continues until the mutation energy upper-bound is reached. Each seed is assigned with a priority score which measures its ability to detect new branches after mutations.

\begin{algorithm}[t]
  \setstretch{0.88}
  \small

    \caption{ \textsc{Branch Searching} }
    \label{alg2}
    
    \KwIn{ Program $\mathcal{P}$, Test case $case$, Vulnerable statements $\mathcal{T}$ }
    \KwOut{ $B_{rare}$ and $B_{vulnerable}$ }
    $b_r \leftarrow FuzzRun(\mathcal{P}, case)$\; 
    $B_{rare} \leftarrow \varnothing$;     \quad\quad\quad\   \tcp*[h]{\emph{Rare Branches}} \\
    $B_{vulnerable} \leftarrow \varnothing$;   \quad   \tcp*[h]{\emph{Vulnerable Branches}}\\

    $\mathcal{R} \leftarrow 0$\;
    $i \leftarrow 0$\;
    \While{$i < |b_r|$}{
    \If{$IsConditionInstruction(i, C_b)$}{
    $\mathcal{R} \leftarrow \mathcal{R}+1$,
    $b_{pre} \leftarrow b_r[0...i+1]$\;
    $c,state \leftarrow StateInference(b_{pre})$\;
    \If{$VulnerableStatementReached(\mathcal{P}, state, c, \mathcal{T})$}{
        $B_{vulnerable}.ADD(b_r)$\;
    }
    }
    $i \leftarrow i+1$\;
    }
    \If{$\mathcal{R} \geqslant 2$}{
        $B_{rare}.ADD(b_r)$\;
    }
   
\end{algorithm}

\subsection{Energy Allocation} 
\label{energy_allocation}
In this subsection, we introduce how IR-Fuzz performs fuzzing energy allocation. Recall that most existing fuzzers treat program branches equally, ignoring the fact that vulnerable code usually takes a tiny fraction of the entire code~\cite{shin2013can,li2020v}. As a result, conventional fuzzers may waste massive resources in fuzzing normal branches instead of rare branches and branches that are more likely to possess bugs. To tackle this problem, we design an energy allocation mechanism, guiding IR-Fuzz to assign fuzzing resources towards these important branches. Specifically, this mechanism consists of two modules: \emph{branch analysis} and \emph{energy schedule}.

\textbf{Branch Analysis.}\quad {We remark that the first challenge is how to pick out the important branches.} To address this, we introduce a branch search algorithm to analyze all branches discovered during fuzzing and focus on two types of branches: \emph{rare} and \emph{vulnerable}, which are defined as follows.

\vspace{3px}
\textbf{Definition 1} (\textbf{Branch):}\quad \emph{Given a path ${p}$ exercised by a test case in smart contract $\mathcal{S}$, we say that $b_r$ is a prefix subpath of $p$ if $b_r$ is a subpath of $p$ and $b_r$ begins with the same starting point as $p$. Further, $b_r$ is a branch  of $p$ if $b_r$ is a  prefix subpath of $p$ and $b_r$ ends with the if-branch or then-branch of a conditional or recurrent statement ({e.g.}, if, require, for, while) in $\mathcal{S}$.}

\textbf{Definition 2} (\textbf{Rare Branch):}\quad \emph{We consider a branch $b_r$ is a rare branch if $b_r$ contains at least two nested conditional statements ({e.g}, two nested \emph{if}). Each rare branch is associated with a {rarity factor} $\mathcal{R}$, which is set to the number of nested conditional statements (at the end of this branch).}

\textbf{Definition 3} (\textbf{Vulnerable Branch):}\quad \emph{Given a branch $b_r$ and a set of vulnerable statements $\mathcal{T}$ that may introduce bugs  (e.g., block.number and call.value), we say that $b_r$ is a \emph{vulnerable} branch when $b_r$ contains a vulnerable statement $t \in \mathcal{T}$. }
\vspace{3px}

\begin{figure}
    \centering
    \includegraphics[width=8.7cm]{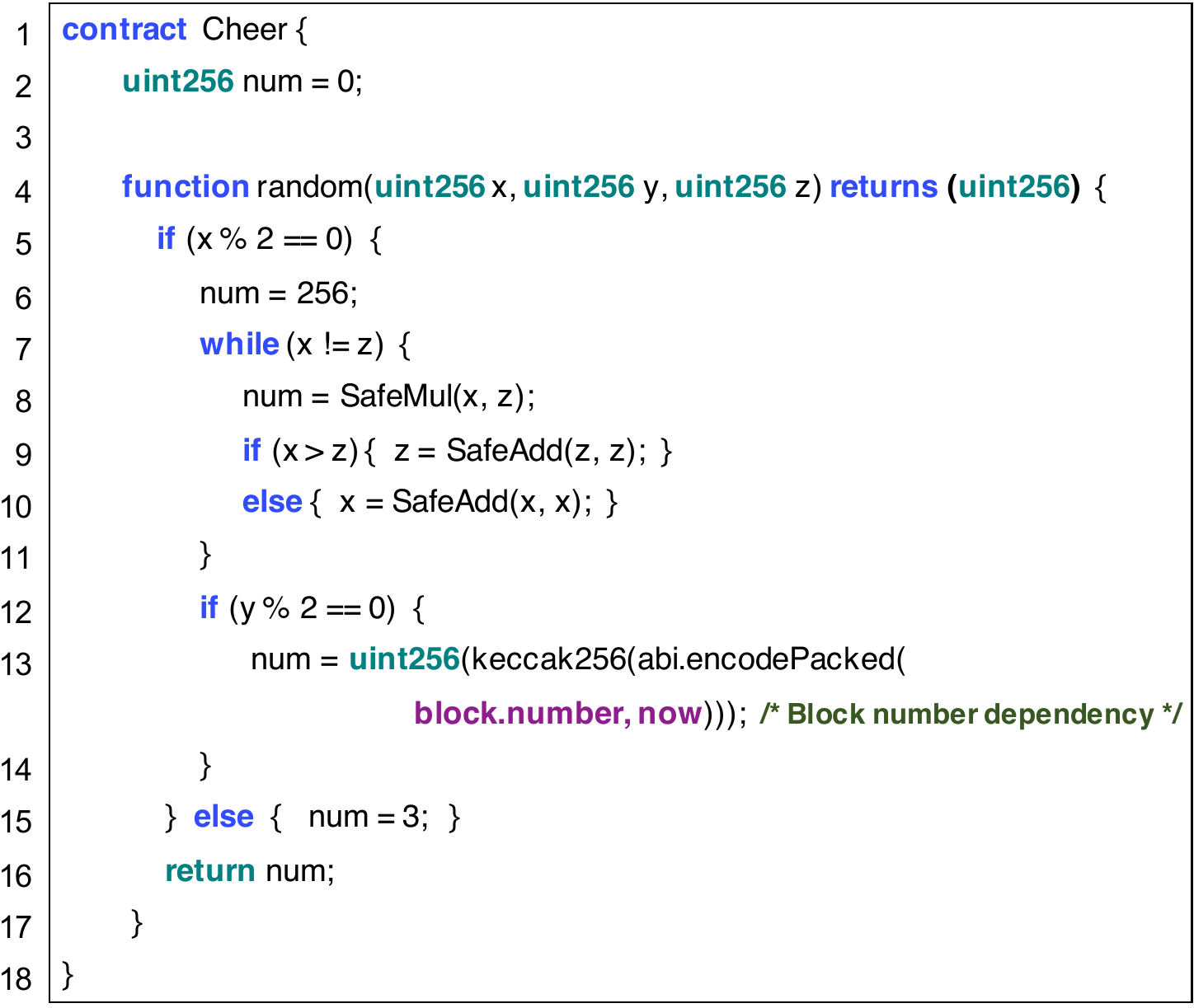}
    \caption{An example contract for illustrating how IR-Fuzz allocates energy to the target branches flexibly.} 
    \label{fig3}
    \vspace{-0.8em}
\end{figure}

\begin{figure*}
    \centering
    \includegraphics[width=18.1cm]{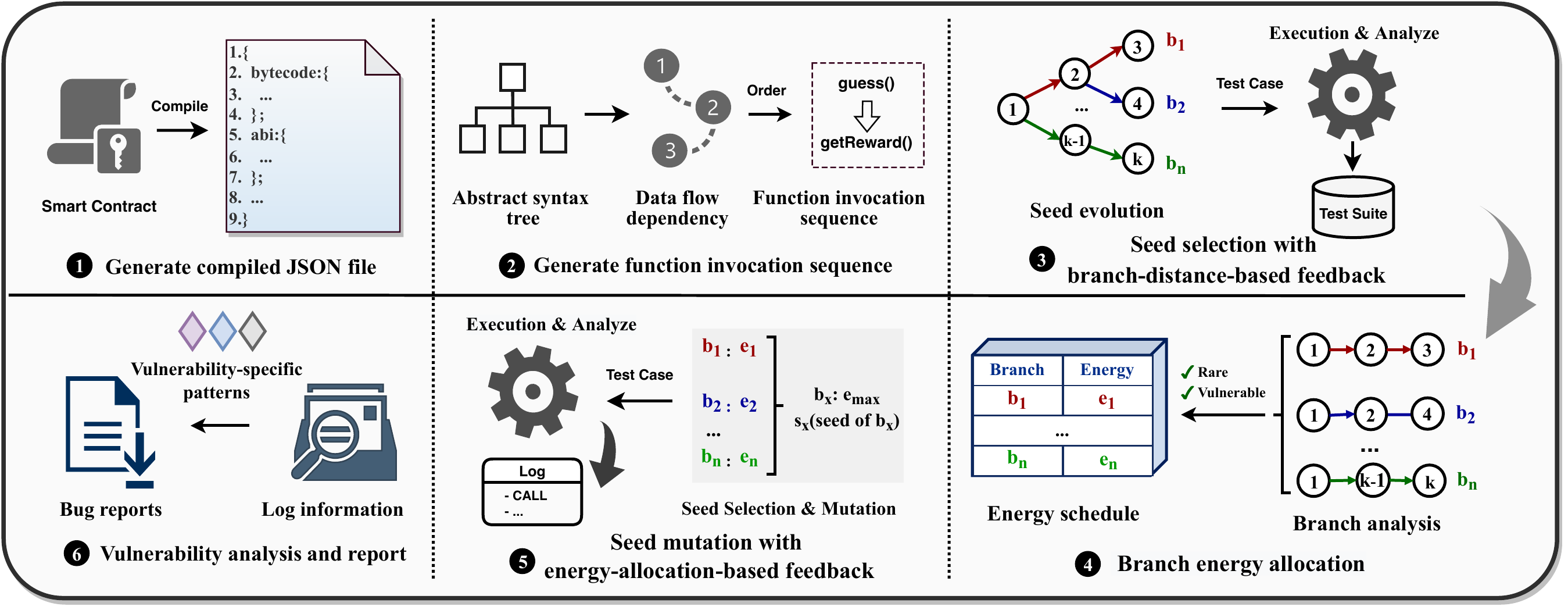}
    \caption{ The workflow of {IR-Fuzz} to reveal the reentrancy vulnerability in the real-world contract of Fig.~\ref{fig1}. }
    \label{workflow}
    \vspace{-0.8em}
\end{figure*}

After empirically scrutinizing real-world smart contracts, we found that over 50\% ($65\%$ and $86\%$ in our experiments) of bugs are located in rare and vulnerable branches. To pick out the two types of branches, we design a branch analysis algorithm shown in Algorithm~\ref{alg2}. Technically, {IR-Fuzz} employs the abstract interpreter $\mathcal{A}$ to pick out the important branches. First, $\mathcal{A}$ analyzes all branches $B_r$ discovered during fuzzing (line 1). Then, the loop from lines 6 to 12 checks whether there exists a branch $b_r$ that reaches the vulnerable statement $t \in \mathcal{T}$ and computes the rarity factor $\mathcal{R}$ for each branch. Afterwards, $\mathcal{A}$ adds the branches with $\mathcal{R}\geqslant2$ into $B_{rare}$ (lines 13-14). Finally, $\mathcal{A}$ obtains a set of rare branches $B_{rare}$ and vulnerable branches $B_{vulnerable}$.

\textbf{Energy Schedule.}\quad We remark that the second challenge is how to assign resources to these important branches. Towards this aim, we formulate a customized energy schedule $\Omega$ to manage the fuzzing energy allocation. This schedule adopts two rules for rare and vulnerable branches, respectively.

\vspace{3px}
\textbf{Rule 3: Energy Allocation for Rare Branches.}\quad \emph{Given branches $b_{r1}$ with $\mathcal{R}_1$ and $b_{r2}$ with $\mathcal{R}_2$ in a same path $p$, where $\mathcal{R}_i$ is rarity factor and $\mathcal{R}_1 < \mathcal{R}_2$. To facilitate that rare branches with higher rarity factors are more sufficiently fuzzed, $\Omega $ assigns energy $\mathcal{r}_1 \ast \mathcal{E}$ to $b_{r1}$ and $\mathcal{r}_2 \ast \mathcal{E}$ to $b_{r2}$ where $\mathcal{r}_1 < \mathcal{r}_2$. }

\textbf{Rule 4: Energy Allocation for Vulnerable Branches.}\quad \emph{Given a branch $b_{r1}$ in a path $p_1$ and a branch $b_{r2}$ in a path $p_2$, $\Omega $ assigns $\alpha \ast \mathcal{E}$ energy to $b_{r1}$ ($\alpha > 1$) and $\mathcal{E}$ energy to $b_{r2}$ when $\mathcal{R}_1 = \mathcal{R}_2$ and $b_{r1} \in B_{vulnerable}$. Coefficient $\alpha$ controls the preference degree for vulnerable branches. }
\vspace{3px}

We use the example of Fig.~\ref{fig3} to show how {IR-Fuzz} works with the energy allocation mechanism. Specifically, {IR-Fuzz} first analyzes all discovered branches and picks out $b_{r}$ at line 12 as the \emph{vulnerable} branch since it contains a vulnerable statement (\emph{i.e., block.number}) at line 13. Then, {IR-Fuzz} calculates the rarity factor $\mathcal{R}$ of this branch, where $\mathcal{R}=2$. As a result, {IR-Fuzz} assigns energy $(\alpha + \mathcal{r}) \ast \mathcal{E}$ to branch $b_{r}$, according to \textbf{Rule 3} and \textbf{Rule 4}. In our experiments, an interesting observation is that state-of-the-art tools like \emph{sFuzz} cannot reach the if-branch at line 12 since they waste too much energy in fuzzing the \emph{while} branch  at line 7. In contrast, {IR-Fuzz} successfully covers this branch and exposes the \emph{block number dependency} vulnerability at line 13 in 8s on average (between 4s and 12s in 5 runs). 

Moreover, {IR-Fuzz} also utilizes the feedback of energy allocation to guide seed mutation. For example, the test cases which cover the vulnerable branches will be selected and fuzzed first. Inspiringly, with the energy allocation mechanism, {IR-Fuzz} can flexibly assign fuzzing resources to the rare and vulnerable branches, increasing overall fuzzing efficiency and branch coverage.

\subsection{Vulnerability Analysis and Report} 
In this stage, {IR-Fuzz} turns to vulnerability analysis and report generation, which reveals vulnerabilities in smart contracts and generates a detailed bug report for further manual confirmation. In particular, we investigate previous works (\emph{e.g.,}~\cite{contractfuzzer,liu2021smart,liu2021combining}) and define the specific patterns for {eight} types of vulnerabilities, namely \emph{timestamp dependency, block number dependency, dangerous delegatecall, Ether frozen, unchecked external call, reentrancy, integer overflow,} and \emph{dangerous Ether strict equality}. We have implemented a pattern analyzer to handle these patterns. {IR-Fuzz} analyzes the fuzzing results and reveals bugs with the assistance of the pattern analyzer. In the following, we show an example of how {IR-Fuzz} exposes a reentrancy vulnerability using patterns. 

\textbf{Vulnerability Pattern Example.}\quad Reentrancy vulnerability is considered as an invocation to \emph{call.value} that can call back to itself through a chain of calls. That is, the invocation of \emph{call.value} is successfully re-entered to perform unexpected repeat money transfers. Specifically, we design two patterns to expose the reentrancy vulnerability. The first pattern \texttt{CALLValueInvocation} checks if there exists an invocation to \emph{call.value} in the contract. The second pattern \texttt{RepeatedCallValue} concerns whether a specific function with \emph{call.value} invocation is called repeatedly during fuzzing. {IR-Fuzz} reports that a function has a reentrancy vulnerability if it fulfills the combined pattern: $ \texttt{CALLValueInvocation}\ \wedge\ \texttt{RepeatedCallValue} $.

\subsection{{IR-Fuzz} Workflow Illustration with an Example}
In this subsection, we take the contract of Fig.~\ref{fig1} as an example to show the workflow of {IR-Fuzz} on revealing a reentrancy vulnerability. The workflow consists of six steps, illustrated in Fig.~\ref{workflow}. Given the contract with source code as input, {IR-Fuzz} first compiles the source code to the JSON file, which consists of EVM bytecode and application binary interface (ABI) (\ding{182} in Fig.~\ref{workflow}). Second, {IR-Fuzz} extracts the abstract syntax tree and captures the data flow dependencies of global variables. By analyzing these dependencies, {IR-Fuzz} infers the function invocation sequence as: \emph{guess()} $\rightarrow$ \emph{getReward()} (\ding{183} in Fig.~\ref{workflow}). Thirdly, {IR-Fuzz} generates test cases for the two function calls and adds high-quality test cases into the test suite based on the {branch distance}-based measure. As such, {IR-Fuzz} effectively generates test cases to cover new branches (\ding{184} in Fig.~\ref{workflow}). Furthermore, {IR-Fuzz} performs the branch analysis to pick out the important branches. Then, it customizes an energy schedule to assign fuzzing resources (\ding{185} in Fig.~\ref{workflow}). {IR-Fuzz} utilizes the feedback of energy allocation to guide seed selection and mutation. After several rounds of fuzzing, {IR-Fuzz} reaches the then-branch at line 18 and triggers the execution of the transfer function (\emph{i.e.,} \emph{call.value}). We develop an attack contract generator, which simulates an \emph{attack} contract that calls the transfer function \emph{getReward()} again. {IR-Fuzz} records the instructions into a log file (\ding{186} in Fig.~\ref{workflow}). Finally, {IR-Fuzz} analyzes the log and determines whether the \emph{call.value} was called multiple times, exposing the reentrancy vulnerability with the assistance of vulnerability-specific patterns (\ding{187} in Fig.~\ref{workflow}). Besides fuzzing the ordered invocation sequence, {IR-Fuzz} further prolongs the sequence to explore other complex states.

\section{Experiments}
In this section, we conduct extensive experiments to evaluate {IR-Fuzz}, seeking to address the following research questions.
\begin{itemize}[noitemsep,wide=0pt, topsep=1pt, leftmargin=\dimexpr\labelwidth + 2\labelsep\relax]
\item \textbf{RQ1:} Can {IR-Fuzz} effectively detect contract vulnerabilities? How is its performance against state-of-the-art tools? 
\item \textbf{RQ2:} Does {IR-Fuzz} achieve higher branch coverage than existing methods?
\item \textbf{RQ3:} How efficient is {IR-Fuzz} in fuzzing smart contracts and generating test cases compared with other fuzzers?
\item \textbf{RQ4:} How much do different components of IR-Fuzz contribute to its performance in branch coverage and vulnerability detection accuracy? 
\end{itemize}
We first introduce the experimental settings, then proceed to answer the above questions. We also present a case study to allow for a better understanding of the proposed approach.

\renewcommand\arraystretch{1.1}
\begin{table}
\centering
    \caption{Summary of vulnerability types supported by state-of-the-art methods. TP is short for timestamp dependency; BN represents block number dependency; DG represents dangerous delegatecall; EF represents Ether frozen; UC represents unchecked external call; RE represents reentrancy; OF represents integer overflow; SE represents dangerous Ether strict equality.} 
    \resizebox{0.495\textwidth}{!}{
        \begin{tabular}{| c | c c c c c c c c | c | c |}
            \hline
            \multirow{2}*{ \textbf{Methods} }    & \multicolumn{8}{c|}{ \textbf{Vulnerability Type} }  & \multirow{2}*{\textbf{\#Citation or \#GitHub Stars}}  &  \multirow{2}*{\textbf{Publication}}	\\	
            \cline{2-9}
           & \textbf{TP}	& \textbf{BN}	& \textbf{DG}	& \textbf{EF}	& \textbf{UC}	& \textbf{RE}	& \textbf{OF}	& \textbf{SE} & &   \\
            \hline          
            Oyente~\cite{oyente}     & $\checkmark$            &             &             &             &             & $\checkmark$            &  $\checkmark$           &        &  1,780 citations  &   CCS'16    \\
            Osiris~\cite{torres2018osiris}           	& $\checkmark$            &             &             &             &             & $\checkmark$            & $\checkmark$            &       & 182 citations   &   ACSAC'18    \\
            Securify~\cite{securify}        	&             &             &             & $\checkmark$            & $\checkmark$            & $\checkmark$            &             &        &  604 citations  &   CCS'18    \\
            ILF~\cite{he2019learning}     		&             & $\checkmark$            & $\checkmark$            & $\checkmark$            & $\checkmark$            &             &             &        &  105 citations  &   CCS'19  \\
                        sFuzz~\cite{nguyen2020sfuzz}      	& $\checkmark$            & $\checkmark$            & $\checkmark$            & $\checkmark$            & $\checkmark$            & $\checkmark$            & $\checkmark$            &        &   91 citations &    ICSE'20   \\
  Mythril~\cite{mythril}                  &     & $\checkmark$        &      $\checkmark$          &             &     & $\checkmark$            &    $\checkmark$     &      &   2,900 GitHub stars  &  White Paper   \\
            ConFuzzius~\cite{torres2021confuzzius}    	&            &       $\checkmark$       &     $\checkmark$            &      $\checkmark$            &      $\checkmark$        &     $\checkmark$            & $\checkmark$            &       &   45 GitHub stars  &   EuroS\&P'21    \\
            \textbf{IR-Fuzz} 	& \textbf{ $\checkmark$ } & \textbf{ $\checkmark$ } & \textbf{ $\checkmark$ } & \textbf{ $\checkmark$ } & \textbf{ $\checkmark$ } & \textbf{ $\checkmark$ } & \textbf{ $\checkmark$ } & \textbf{ $\checkmark$ } &  -- &  --  \\
            \hline
        \end{tabular}}
        \label{baselines}
        \vspace{-0.8em}
\end{table}

\subsection{Experimental Setup}
\label{experimental_setup}

\textbf{Implementation.}\quad {IR-Fuzz} in total contains 9K+ lines of C++ code, {which is released for public use at {\url{https://github.com/Messi-Q/IR-Fuzz}} }. We implemented {IR-Fuzz} on the basis of sFuzz~\cite{nguyen2020sfuzz} (a state-of-the-art smart contract fuzzer). 

\textbf{Baselines.}\quad {In the experiments, we include seven open-source methods that either have a high number of citations or receive many stars in GitHub. The methods are summarized in Table~\ref{baselines}, where we illustrate the vulnerability types that they can detect, their numbers of citations or GitHub stars, and their publication information.} For fuzzing tools, we select ConFuzzius~\cite{torres2021confuzzius}, ILF~\cite{he2019learning}, and sFuzz~\cite{nguyen2020sfuzz}, which achieve state-of-the-art performance and support at least four vulnerability types on smart contracts.  For static analysis tools, we select Mythril~\cite{mythril}, Oyente~\cite{oyente}, Osiris~\cite{torres2018osiris}, and Securify~\cite{securify}, which are well-known vulnerability checkers for smart contracts. We compare {IR-Fuzz} with them in terms of branch coverage, effectiveness, and efficiency. All experiments are conducted on a computer equipped with an Intel Core i9 CPU at 3.3GHz, a GPU at 2080Ti, and 64GB Memory. Each experiment is repeated ten times, we report the average results.  

\textbf{Dataset.}\quad We obtain the dataset by crawling Etherscan~\cite{Etherscan} verified contracts, which are real-world smart contracts deployed on Ethereum Mainnet. In practice, we removed 5,074 duplicate contracts by comparing the hash of the contract binary code. Our final dataset contains a total 12,515 smart contacts that have source code. As listed in Table~\ref{baselines}, we focus on {eight types of vulnerabilities} in the dataset, \emph{namely} timestamp dependency (TP), block number dependency (BN), dangerous delegatecall (DG), Ether frozen (EF), unchecked external call (UC), reentrancy (RE), integer overflow (OF), and dangerous Ether strict equality (SE). We deployed all smart contacts of the dataset to a local Ethereum test network for experiments. For the ground truth labels of smart contracts,  we define vulnerability-specific patterns for each kind of vulnerability to give a preliminary label and then manually check whether a smart contract in the dataset indeed has a certain vulnerability. In particular, using the defined vulnerability-specific patterns (\emph{e.g., keyword matching}), we could find smart contracts that may have vulnerabilities and save our time on labeling those contracts that are safe (\emph{e.g., a contract with no `{call.value}' invocation will not have reentrancy vulnerabilities}).

\subsection{Effectiveness (RQ1)} 
First, we benchmark IR-Fuzz against existing vulnerability detection methods. We count the number of smart contracts that have vulnerabilities and are identified by each method, and present the accuracy, true positives, and false positives of each method. 

\textbf{Comparing IR-Fuzz to State-of-the-arts.}\quad We first compare {IR-Fuzz} to other fuzzers and existing static analysis tools. Quantitative experimental results of each method are summarized in Table~\ref{accuracy}. From the table, we obtain the following observations. 
\textbf{(1)} Compared with other methods, {IR-Fuzz} is able to identify more vulnerabilities. Inspiringly, {IR-Fuzz} has achieved a high accuracy (more than 90\%) on all eight types of vulnerabilities. 
\textbf{(2)} {IR-Fuzz} consistently outperforms state-of-the-art methods by a large margin in detecting each type of vulnerability. For example, for Ether frozen vulnerability (EF), {IR-Fuzz} gains 15.63\% and 16.14\% accuracy improvements over  \emph{Securify} and \emph{ConFuzzius}. These strong empirical evidences suggest the great potential of {IR-Fuzz} to identify smart contract vulnerabilities. We attribute its superior performance to the key modules proposed, \emph{i.e.,} sequence generation, seed optimization, and energy allocation, which boost the capability of IR-Fuzz in improving branch coverage and hunting vulnerabilities. 
\textbf{(3)} Promisingly, {IR-Fuzz} discovers a new kind of smart contract vulnerability, \emph{i.e.,} dangerous Ether strict equality (SE). To the best of our knowledge, this vulnerability cannot yet be detected by current automatic tools. We also present an illustrative case study on how our method detects this vulnerability in \S\ref{case_study}.

\renewcommand\arraystretch{1.0}
\begin{table}
  \centering
  \caption{Accuracy comparison (\%) on different methods, including static analysis tools, fuzzers, and {IR-Fuzz}. `n/a' denotes that a tool cannot detect the specific vulnerability. }
\resizebox{0.49\textwidth}{!}{
\begin{tabular}{|c|cccccccc|}
\hline
\multirow{2}{*}{\textbf{Methods}}&\multicolumn{8}{c|}{\textbf{Vulnerability Type (Accuracy)}}  \\
 \cline{2-9}
 & \textbf{TP} & \textbf{BN} & \textbf{DG} & \textbf{FE} & \textbf{UC} & \textbf{RE} & \textbf{OF} & \textbf{SE}  \\ 
 \hline
Mythril \cite{mythril}    & n/a               & 89.97               & 70.95                   & n/a                     & n/a                     & 95.09               & 89.87             & n/a                       \\
Oyente \cite{oyente}     & 86.86            & n/a                  & n/a                   & n/a                     & n/a                     & 94.61               & 74.76             & n/a                       \\
Osiris \cite{torres2018osiris}     & 86.56            & n/a                  & n/a                   & n/a                     & n/a                     & 93.28               & 74.80            & n/a                       \\
Securify \cite{securify}   & n/a               & n/a                  & n/a                   & 79.42                   & 91.24                  & 91.52               & n/a               & n/a                       \\
ILF    \cite{he2019learning}    & n/a               & 87.53                & 80.99                 & 78.65                   & 94.71                   & n/a                 & n/a               & n/a                       \\
sFuzz \cite{nguyen2020sfuzz}    & 87.25             & 88.37               &83.33               & 83.85                   & 94.26                   & 95.20               & 89.98             & n/a                       \\
ConFuzzius \cite{torres2021confuzzius}  & n/a               & 87.70                & 80.47                 & 78.91                   & 94.68                  & 93.33               & 77.35             & n/a                       \\ 
\textbf{IR-Fuzz}   & \textbf{90.25}    & \textbf{94.18}       & \textbf{95.33}        & \textbf{95.05}          & \textbf{98.10}          & \textbf{98.77}      & \textbf{98.79}    & \textbf{99.73}    \\ 
\hline      
\end{tabular}
}
\label{accuracy}
\end{table}

\renewcommand\arraystretch{1.0}
\begin{table}
\centering
\caption{True and false positives of each method in identifying the eight types of smart contract vulnerabilities.} 

\resizebox{0.49\textwidth}{!}{
\begin{tabular}{|c|cccccccc|c|}
\hline
\multirow{2}{*}{\textbf{Methods}} &\multicolumn{8}{c|}{\textbf{Vulnerability Type (True / False Positives)}}  & \multirow{2}{*}{\textbf{Total}} \\ 
 \cline{2-9}
&\multicolumn{1}{c}{\textbf{TP}} & \multicolumn{1}{c}{\textbf{BN}} & \multicolumn{1}{c}{\textbf{DG}} & \multicolumn{1}{c}{\textbf{FE}} & \multicolumn{1}{c}{\textbf{UC}} & \multicolumn{1}{c}{\textbf{RE}} & \multicolumn{1}{c}{\textbf{OF}} &\textbf{SE} &  \\ 
 \hline
Mythril   \cite{mythril}             & n/a                       & 4/63                     & 20/20                       & n/a                       & n/a                        & 0/62                    & 10/245                  & n/a   &           34             \\  
Oyente     \cite{oyente}            & 12/6                     & n/a                        & n/a                        & n/a        & n/a                   & 8/87                        & 16/637                                         & n/a   &            36            \\
Osiris    \cite{torres2018osiris}              & 4/5                    & n/a                        & n/a             & n/a           & n/a                        & 12/139                        & 12/632                                           & n/a   &     28                   \\ 
Securify    \cite{securify}           & n/a                       & n/a                        & n/a                        & 0/0                     & 7/208                   & 4/194                   & n/a                        & n/a   &           11             \\ 
ILF       \cite{he2019learning}             & n/a                        & 0/103                     & 8/4                     & 0/3                     & 5/82                    & n/a                       &   n/a               & n/a   &                13        \\ 
sFuzz     \cite{nguyen2020sfuzz}             & 23/8                    & 20/108                    & 20/7                    & 20/3                    & 7/100                    & 10/68                    & 3/235                   & n/a   &              103          \\ 
ConFuzzius   \cite{torres2021confuzzius}          & n/a                        & 20/120                   & 4/2                     & 0/2                     & 8/86                    & 6/131                    &     10/565                      & n/a   &               48         \\ 
\textbf{IR-Fuzz}              & \textbf{92/5}                    & \textbf{26/3}                    & \textbf{58/0}                    & \textbf{65/0}                    &\textbf{ 83/36  }                  & \textbf{95/20}                    & \textbf{21/10}                    & \textbf{45/0} &           \textbf{485}             \\ 
 \cline{1-10}
\hline
\end{tabular}
}
 \label{fp_fn}
 \vspace{-0.8em}
\end{table}

\textbf{Analysis of True and False Positives.}\quad 
To further evaluate the effectiveness of {IR-Fuzz}, we examine the identified vulnerable contracts to see whether they are true positives or not. Table~\ref{fp_fn} demonstrates the number of vulnerable contracts discovered by each method, as well as the numbers of true positives and false positives of each method. 
(1) From Table~\ref{fp_fn}, we observe that existing methods have not yet obtained a high true positive rate on the eight types of vulnerabilities. For example, for unchecked external call vulnerability (UC), \emph{Securify} and \emph{sFuzz} generate 7 true positives, while \emph{ConFuzzius} and \emph{ILF} obtain 8 and 2 true positives, respectively. This is mainly due to the reason that conventional tools ignore handling exceptions for the return value of external calls. 
(2) Moreover, we also find that existing methods have high false positives. For block number dependency vulnerability (BN), fuzzing tools \emph{sFuzz}, \emph{ConFuzzius}, and \emph{ILF} produce over 100 false positives. For integer overflow vulnerability (OF), 632, 637, and 565 false positives are reported by \emph{Osiris}, \emph{Oyente}, and \emph{ConFuzzius}, respectively. The high false positives of these methods may stem from two facts: {(i)} Most methods tend to detect vulnerabilities using a few simple but imprecise patterns, \emph{e.g.}, identifying block number vulnerability by crudely checking whether there is a \emph{block.number} statement in the function; {(ii)} Many tools conservatively assume that all arithmetic operations returning a negative value are  vulnerable, resulting in high false positives.

{IR-Fuzz} reports more true positives than other methods. For example, for timestamp dependency vulnerability (TP), {IR-Fuzz} generates 92 true positives, 88, 80, and 69 more than \emph{Osiris}, \emph{Oyente}, and \emph{sFuzz}, respectively. In total, {IR-Fuzz} finds vulnerabilities in 485 contracts, roughly 4.7 times more than \emph{sFuzz}, which ranks the second. For reentrancy vulnerability (RE), {IR-Fuzz} produces 95 true positives, which significantly outperforms the state-of-the-art tool \emph{Osiris}. More importantly, {IR-Fuzz} can precisely detect a new kind of vulnerability (SE) without reporting any false positives. We attribute the good performance of {IR-Fuzz} to the fact that it integrates the three presented new techniques, which are able to supplement each other for precise bug detection. In summary, {IR-Fuzz} can effectively identify various vulnerabilities in smart contracts, surpassing existing static analysis tools and fuzzers by a large margin.

\begin{figure}
\centering
\subfigure[Branch coverage of different methods on small contracts]{
\includegraphics[width=4.12cm]{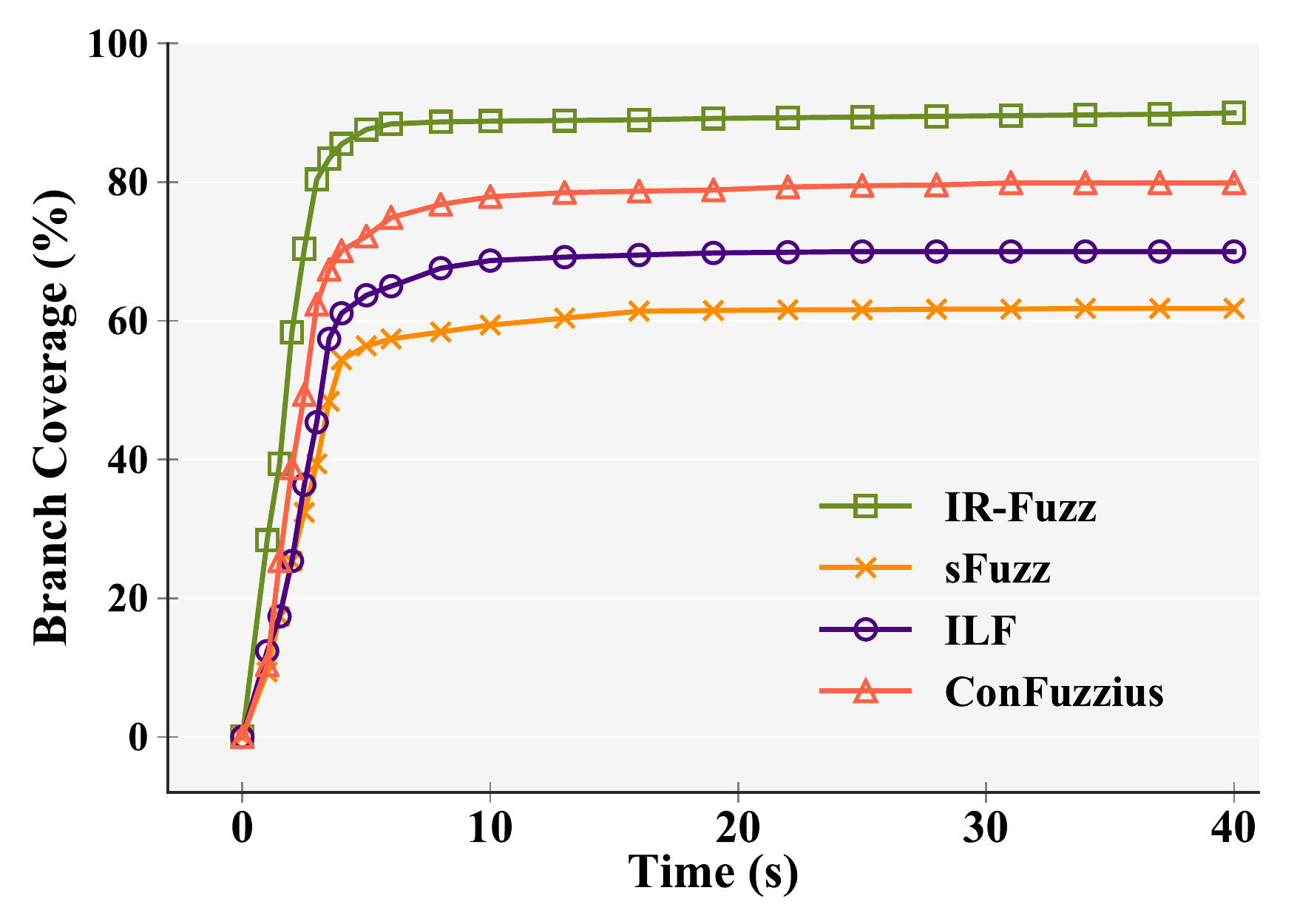}
}
\subfigure[Branch coverage of different methods  on large contracts]{
\includegraphics[width=4.12cm]{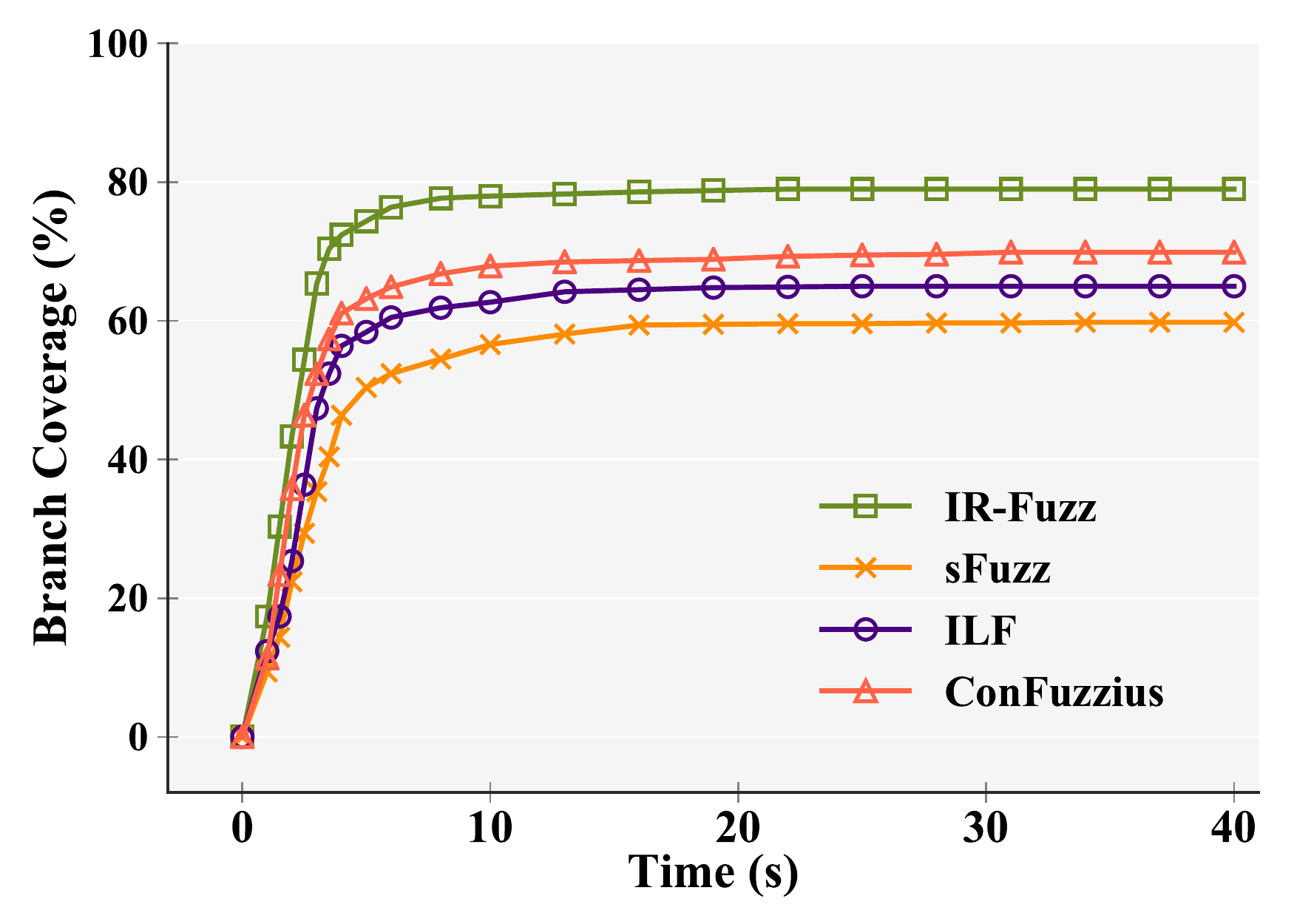}
}
\caption{Curves comparison: the tendency of branch coverage over time on different fuzzers.}
\label{coverage_time}
\vspace{-0.8em}
\end{figure}

\subsection{Branch Coverage (RQ2)} 
We now present evaluation results on branch coverage of {IR-Fuzz}. We measure the number of distinct branches covered by the generated test cases in the test suite. Moreover, to examine the branch coverage on contracts with different sizes, we follow the settings of previous work~\cite{he2019learning} and split the dataset into 1,885 large contracts ($\geq$3,600 instructions) and 10,630 small ones ($<$3,600 instructions).

We compare with other fuzzers (\emph{i.e.,} sFuzz, ILF, and ConFuzzius). Particularly, we visualize the comparison results on small contracts in Fig.~\ref{coverage_time}(a) and on large contracts in Fig.~\ref{coverage_time}(b), respectively. We plot the tendency of branch coverage over time. It can be seen that {IR-Fuzz} consistently outperforms other fuzzers. Quantitatively, {IR-Fuzz} achieves 90.10\% coverage on small contracts, 28.20\%, 20.10\%, and 10.10\% higher than sFuzz, ILF, and ConFuzzius, respectively. On large contracts, {IR-Fuzz} achieves 19.20\%, 14.00\%, and 9.10\% higher coverage, respectively. Moreover, we also observe that {IR-Fuzz} reaches the highest coverage with less time required than other fuzzers. On average, {IR-Fuzz} spent only 10s to achieve the highest coverage (\emph{i.e.,} 90.10\% on small contracts and 79.10\% on large contracts), while the other three fuzzers spent 18s, 16s, 13s, respectively. 

We conjecture that the advantages of IR-Fuzz in achieving high branch coverage come from three aspects. \emph{First,} IR-Fuzz generates the high-quality function invocation sequence by adopting a dependency-aware sequence generation strategy, enforcing the fuzzer to tap into richer states. \emph{Second,} IR-Fuzz employs a branch distance-based measure to iteratively optimize the generated test cases, steering fuzzing towards covering new branches. \emph{Thirdly,} IR-Fuzz takes into account the significance of rare branches and branches that are likely to have vulnerabilities, and designs an energy allocation mechanism to flexibly guide fuzzing energy allocation towards these critical branches. Moreover, IR-Fuzz utilizes the feedback results generated by the energy allocation mechanism to guide seed mutation, which further increases branch coverage.

\begin{figure}
\centering
\subfigure[Average execution time]{
\includegraphics[width=4.15cm]{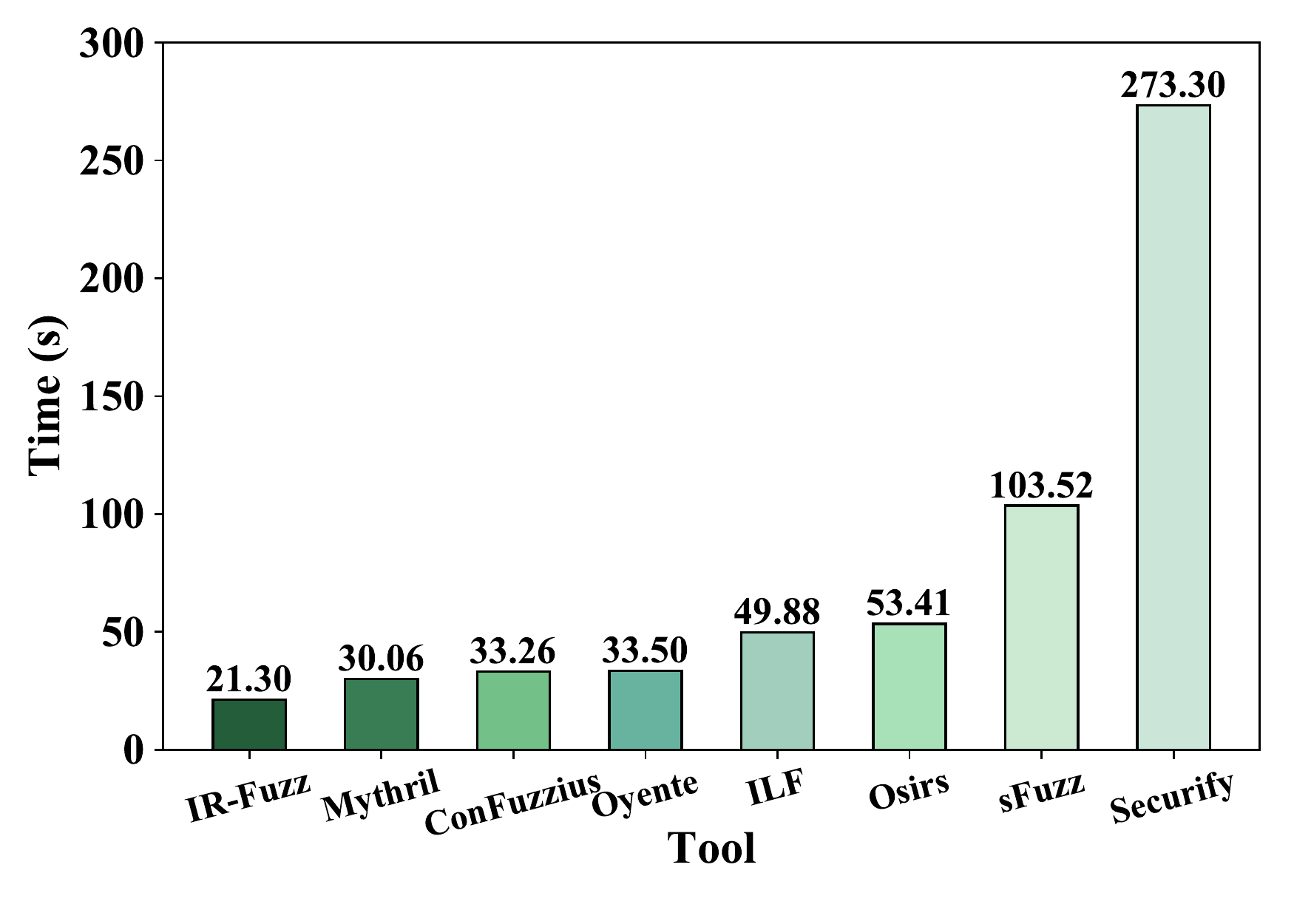}
}
\subfigure[Number of generated test cases]{
\includegraphics[width=3.95cm]{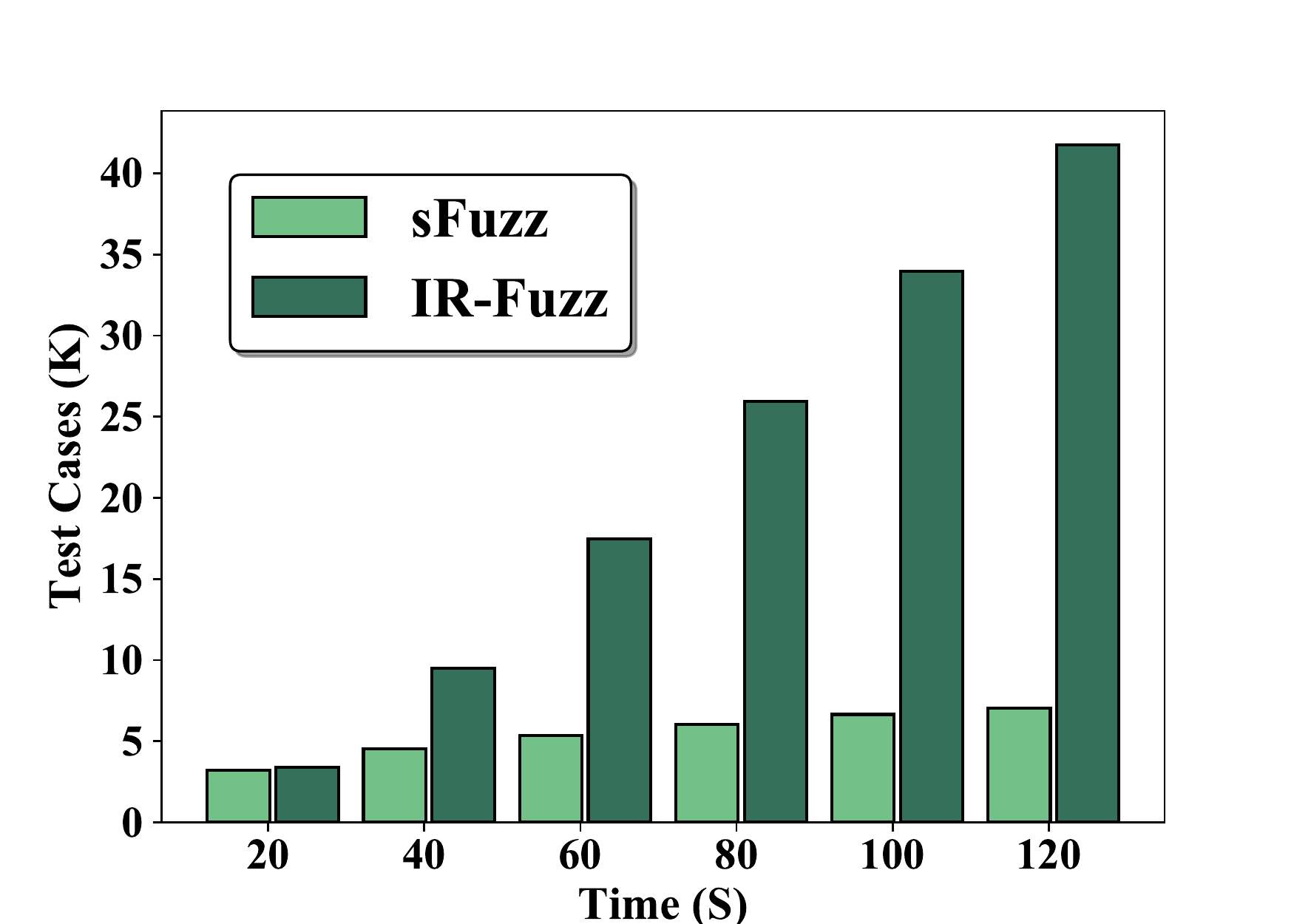}
}
\caption{Visual comparison of efficiency on different tools.}
\label{time_tools}
\vspace{-0.8em}
\end{figure}

\subsection{Efficiency (RQ3)} 
In this subsection, we systematically examine the efficiency of {IR-Fuzz} and compare it against other methods.   

\emph{First,} we conduct experiments to measure the overhead of {IR-Fuzz} by calculating the average execution time on each contract. We run {IR-Fuzz} on the whole dataset, revealing that it spends 21.30s per contract on average. Fig.~\ref{time_tools}(a) compares {IR-Fuzz} to other methods in terms of the average execution time. From the figure, we observe that IR-Fuzz is significantly more efficient than others. Particularly, its average execution time is 251s and 82.22s faster than \emph{Securify} and \emph{sFuzz}, respectively. We believe the reasons for the much faster speed of IR-Fuzz are as follows. \textbf{(1)} {IR-Fuzz} can quickly infer the ordered function invocation sequence, accelerating fuzzing execution. \textbf{(2)} {IR-Fuzz} adopts the {branch distance}-based measure to boost its efficiency in generating test cases, which requires much fewer mutations to reach a target branch. \textbf{(3)} {IR-Fuzz} leverages the energy allocation mechanism to flexibly assign fuzzing resources, saving overall fuzzing time.

\emph{Next,} we further measure the efficiency of {IR-Fuzz} by counting how many test cases are generated over time. Specifically, each contract is run for 120 seconds in the experiment. We show the average statistics in Fig.~\ref{time_tools}(b), where the $x$-axis represents how long a contract is fuzzed, and the $y$-axis denotes the number of test cases generated during fuzzing. From Fig.~\ref{time_tools}(b), we can learn that \textbf{(1)} {IR-Fuzz} significantly generates more test cases than \emph{sFuzz} within the same time interval. On average, {IR-Fuzz} generates approximately 350 test cases per second, 290 more than \emph{sFuzz}; \textbf{(2)} The number of test cases generated by IR-Fuzz has increased rapidly over time while the process is slow in \emph{sFuzz}. These evidences reveal that IR-Fuzz can efficiently generate test cases for fuzzing smart contracts.

\subsection{Ablation Study (RQ4)}
\label{ablation_study}
By default, {IR-Fuzz} adopts the proposed \emph{sequence generation} strategy to generate the function invocation sequence. It is interesting to see the effect of removing this strategy. Moreover, {IR-Fuzz} utilizes a \emph{branch distance}-based measure to select and evolve test cases iteratively. We are curious about how much this method contributes to the performance of {IR-Fuzz}. Finally, {IR-Fuzz} introduces an \emph{energy allocation} mechanism to flexibly guide fuzzing resource allocation. It is useful to evaluate the contributions of this mechanism by removing it from {IR-Fuzz} as well. In what follows, we conduct experiments to study the three components, respectively.

\renewcommand\arraystretch{1.2}
\begin{table}
\centering
\caption{Accuracy and coverage comparison (\%) between {IR-Fuzz} and its variants.}
\resizebox{0.495\textwidth}{!}{
\begin{tabular}{|c|cccccccc|c|}
\hline
\multirow{2}{*}{\textbf{Method}} & \multicolumn{8}{c|}{\textbf{Vulnerability Type (Accuracy)}} & \multirow{2}{*}{\textbf{Coverage}}    \\ 
\cline{2-9}
  & \textbf{TP} & \textbf{BN} & \textbf{DG} & \textbf{FE} & \textbf{UC} & \textbf{RE} & \textbf{OF} & \textbf{SE} &     \\ 
\hline
IR-Fuzz-WSG &90.01 &93.38 &94.48 &92.73 &96.70 &94.80 &93.86 &96.42 &62.03 \\
IR-Fuzz-WDM &89.94 &93.56 &93.31 &92.15 &96.26 &94.51 &92.02 &94.12 &69.89 \\
IR-Fuzz-WEA &89.22 &91.00 &91.86 &91.00 &96.04 &92.49 &91.60 &95.82 &42.63 \\
\textbf{IR-Fuzz} &\textbf{90.05} &\textbf{93.79} &\textbf{95.06} &\textbf{94.48} &\textbf{98.03} &\textbf{98.73} &\textbf{98.73} &\textbf{99.73} &\textbf{85.65} \\
\hline   
\end{tabular}
}
\label{tab_ablation}
\vspace{-0.8em}
\end{table}

\textbf{Study on Sequence Generation Strategy.}\quad We removed the sequence generation strategy (introduced in \S\ref{order_sequence}) from {IR-Fuzz} and replaced it with a random sequence construction method. This variant is denoted as {IR-Fuzz-WSG}, where {WSG} is short for without sequence generation strategy. Quantitative results are summarized in Table~\ref{tab_ablation}. We can observe that the performance of {IR-Fuzz} is significantly better than {IR-Fuzz-WSG}. For example, on the reentrancy detection task, {IR-Fuzz} achieves 3.97\% and 23.62\% improvement in terms of accuracy and branch coverage, respectively. 

\textbf{Study on Branch Distance-based Seed Evolution Paradigm.}\quad To evaluate the effect of the {branch distance}-based seed evolution paradigm, we analyze the performance of {IR-Fuzz} with and without it, respectively. Towards this aim, we modify {IR-Fuzz} by removing this mechanism, utilizing random test case generation. This variant is denoted as {IR-Fuzz-WDM}, where {WDM} is short for without the distance measure mechanism. The empirical findings are demonstrated in Table~\ref{tab_ablation}, where we can observe that the accuracy and branch coverage of {IR-Fuzz-WDM} are lower than IR-Fuzz by an average of 2.03\% and 15.76\% on the eight types of vulnerabilities. This reveals that incorporating the {branch distance}-based measure is necessary and critical to improve the performance of {IR-Fuzz}.

\textbf{Study on Energy Allocation Mechanism.}\quad We further investigate the impact of the energy allocation mechanism in {IR-Fuzz}. Specifically, we remove this mechanism while replacing it with assigning fuzzing energy equally to every branch. This new variant is termed as {IR-Fuzz-WEA}, namely {IR-Fuzz} without an energy allocation mechanism. The comparison results are presented in Table~\ref{tab_ablation}, where all eight types of vulnerabilities are involved. We can clearly see that the accuracy and branch coverage of {IR-Fuzz-WEA} are lower than IR-Fuzz by an average of 4.87\% and 43.02\%. This suggests that the energy allocation mechanism contributes to significant performance gains in {IR-Fuzz}.

\begin{figure}
    \centering
    \includegraphics[width=8.7cm]{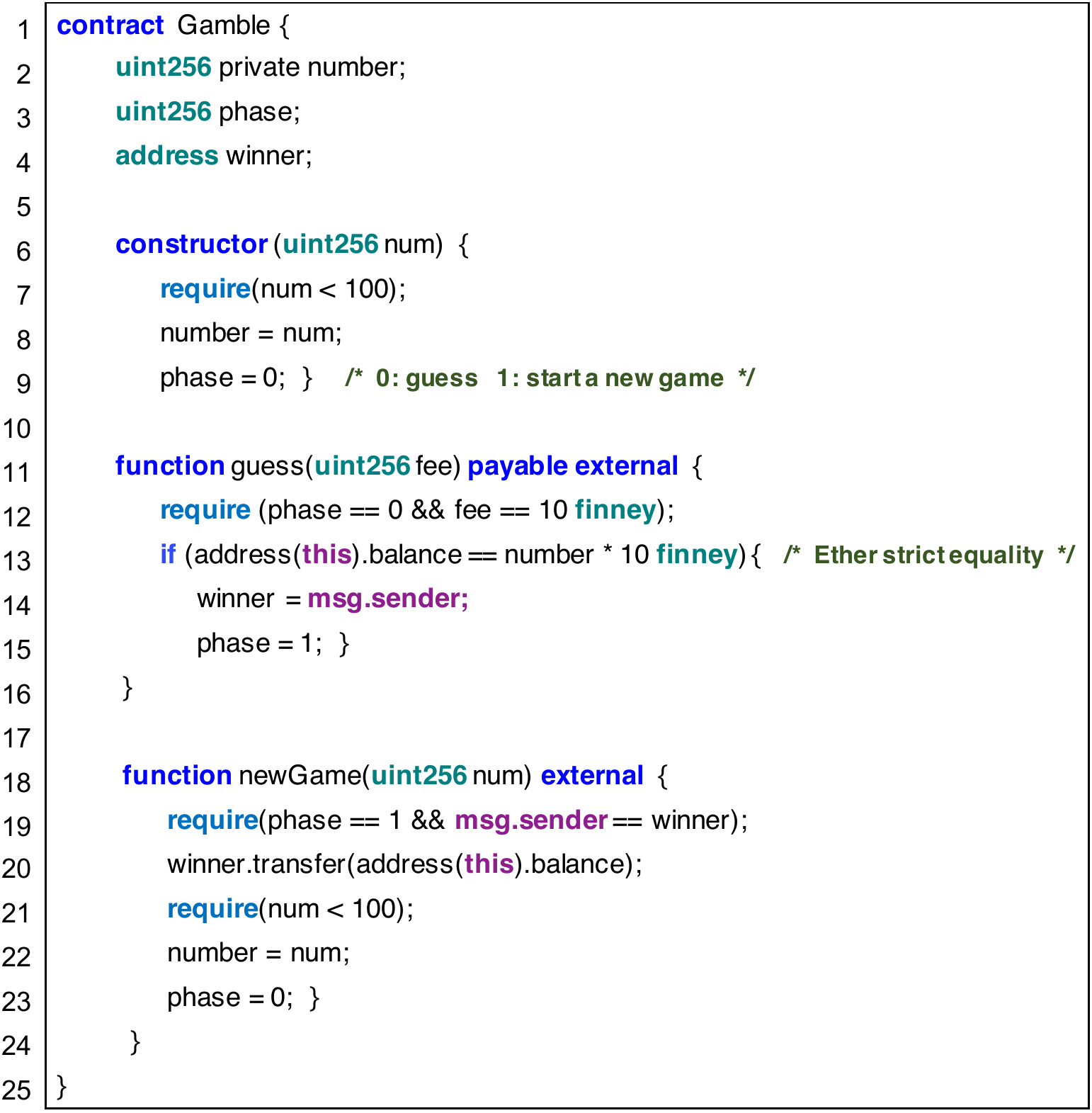}
    \caption{An example contract where {IR-Fuzz} detects a new kind of vulnerability, \emph{i.e.}, dangerous Ether strict equality.}
    \label{fig_guess}
\end{figure}

\subsection{Case Study}
\label{case_study}
We now present a case study on a new vulnerability (\emph{i.e.,} dangerous Ether strict equality). To our knowledge, existing investigated methods cannot expose this vulnerability yet. Fig.~\ref{fig_guess} shows a simplified example that implements a gambling game. A user can join the game by transferring participation fees with 10 finney. If a user is the \emph{number}-th participant, he will become the winner of the game (line 14). The winner can obtain the whole balance of the contract by calling \emph{newGame()} and starting a next round of the game. However, if the contract owner had pre-stored some Ethers in the contract, the balance of the contract will never be equal to the sum of users' participation fees (namely, the branch condition at line 14 will never be satisfied). This indicates that there will be no winner in the game, and the participation fees in the contract will be permanently frozen. 

We empirically checked this contract using existing tools and manually inspected their generated reports. Unfortunately, the dangerous Ether strict equality vulnerability cannot yet be detected by these methods. In contrast, {IR-Fuzz} successfully identifies this vulnerability. Specifically, $T_1$: IR-Fuzz infers the function invocation sequence as: \emph{guess()}$\rightarrow$ \emph{newGame()} and generates a test case to cover the requirement at line 13. $T_2$: Record the instruction {\ttfamily BALANCE} when the fuzzing process reaches line 14. $T_3$: Check if {\ttfamily BALANCE} is followed by the jump or compare instructions. $T_4$: IR-Fuzz finds that line 14 is reachable and the vulnerability-specific patterns of dangerous Ether strict equality are triggered, outputting that the contract has such a vulnerability.

\section{Discussion}
\label{discussion}
In this section, we discuss the limitations of IR-Fuzz and potential future improvements. 

\textbf{Sequence Generation Analysis.}\quad 
IR-Fuzz generates the ordered function invocation sequence with the guidance of the order priority computation rules mentioned in \S\ref{order_sequence}. We calculate the order priority of function calls in the sequence by analyzing the data flow dependencies of global variables. In the case that several functions perform frequent write and read operations on global variables, the calculation of function order priority may bring a certain amount of computation overhead. 

\textbf{Seed Mutation Optimization.}\quad  
IR-Fuzz refers to several seed mutation strategies adopted from AFL, usually using bit manipulation techniques, \emph{e.g.,} bit flipping. However, such a method still bears the problem of generating repetitive and invalid test cases. Moreover, arbitrarily mutating bits of a test input may ignore certain critical parts of the input that should not be mutated, reducing the probability of hitting the branches guarded by strict conditions. Therefore, in the subsequent work, we may focus on enabling the fuzzer not to mutate these crucial parts of a test case, making the fuzzing trigger deep and complex states.

\section{Conclusion}
In this paper, we present IR-Fuzz, a fully automatic fuzzing framework equipped with invocation ordering  and crucial branch revisiting, to detect vulnerabilities in smart contracts. Specifically, we propose a sequence generation strategy consisting of invocation ordering and prolongation to generate the high-quality function invocation sequence, enforcing the fuzzer to trigger complex and deep states. Furthermore, we design a seed optimization paradigm that engages a branch distance-based measure to evolve test cases iteratively towards a target branch, alleviating the randomness of test case generation. Finally, we develop an energy allocation mechanism to flexibly guide fuzzing resource allocation towards rare and vulnerable  branches, improving the overall fuzzing efficiency and branch coverage. Experimental results demonstrate that IR-Fuzz significantly surpasses state-of-the-art fuzzing approaches by a large margin. Our implementation and dataset are released to facilitate future research. The presented techniques in IR-Fuzz might also be transferable to fuzz other software programs.

\footnotesize
\bibliographystyle{IEEEtran}
\bibliography{tifs-template}

\end{sloppypar}

\end{document}